\documentclass[11pt,a4paper]{emulateapj}

\newcommand{\Lbol}{$\hbox{L}_{\rm{bol}}$}
\newcommand{\Lsol}{$\hbox{L}_\odot$}
\newcommand{\Msol}{$\hbox{M}_\odot$}
\newcommand{\Av}{$\hbox{A}_V$}
\newcommand{\HH}{\hbox{${\rm H}_2$}}  

\shorttitle{Physical structure and fragmentation of IM protostellar cores in Orion }
\shortauthors{van Kempen, T.A., et al.}

\usepackage{natbib}

\begin{document}


\title{The small-scale physical structure and fragmentation difference of two embedded intermediate mass protostars in Orion}

\author{T.A. van Kempen}
\affil{Joint ALMA offices, Alonso de Cordova 3107, Vitacura, \\ Santiago, Chile}
\affil{Leiden Observatory, Leiden University,Niels Bohrweg 2, 2333 CA Leiden,  Netherlands}
\affil{Harvard-Smithsonian Center for Astrophysics, 60 Garden Street, Cambridge, \\ MA 02138, United States}
\email{tkempen@alma.cl}
\author{S. N. Longmore}
\affil{European Southern Observatory, Karl-Schwarzschild-Str. 2
85748 Garching bei München
Germany}
\affil{Harvard-Smithsonian Center for Astrophysics, 60 Garden Street, Cambridge, MA 02138, United States}
\author{D. Johnstone}
\affil{National Research Council Canada,  Herzberg Institute for Astronomy, 5071 West Saanich Road, Victoria, British Columbia, Canada}
\affil{Dept. of Physics \& Astronomy,
University of Victoria,
Elliott Building, 3800 Finnerty Rd,
Victoria, BC, V8P 5C2 Canada}
\author{T. Pillai}
\affil{Caltech, MC 249-17
1200 East California Blvd,
Pasadena CA 91125 United States}

\author{A. Fuente}
\affil{Observatorio Astronómico Nacional (OAN), Apdo. 112, 28803 Alcalá de Henares, 
Madrid, Spain}



\begin{abstract}
 Intermediate mass protostars, the bridge between the very common solar-like protostars  and the  more massive, but rarer, O and B stars, can only be studied at high physical spatial resolutions in a handful of clouds.
 In this paper we present and analyze the continuum  results from an observing campaign at the Submillimeter Array targeting two well-studied intermediate mass protostars in Orion, NGC 2071 and L1641 S3 MMS 1. 
The extended SMA (eSMA) probes structure at angular resolutions up to 0.2$"$, revealing protostellar disks on scales of $\sim$200 AU. Continuum flux measurements on these scales indicate that a significant amount of mass, a few tens of \Msol, are present.
 Envelope, stellar, and disk masses are derived using both compact, extended and eSMA configurations and compared against SED-fitting models. We hypothesize that fragmentation into three components occurred within NGC 2071 at an early time, when the envelopes were less than 10$\%$ of their current masses, e.g. $<$ 0.5 \Msol. No fragmentation occurred for L1641 S3 MMS 1. For NGC 2071 evidence is given that the bulk of the envelope material currently around each source was accreted after the initial fragmentation. In addition, about 30$\%$ of the total core mass is not yet associated to one of the three sources. A global accretion model is favored and a potential accretion history of NGC 2071 is presented. It is shown that the relatively low level of fragmentation in NGC 2071 was stifled compared to the expected fragmentation from a Jean's argument. 
\end{abstract}





\def\placeFigureSCUBA{
\begin{figure*}
\begin{center}
\includegraphics[width=500pt]{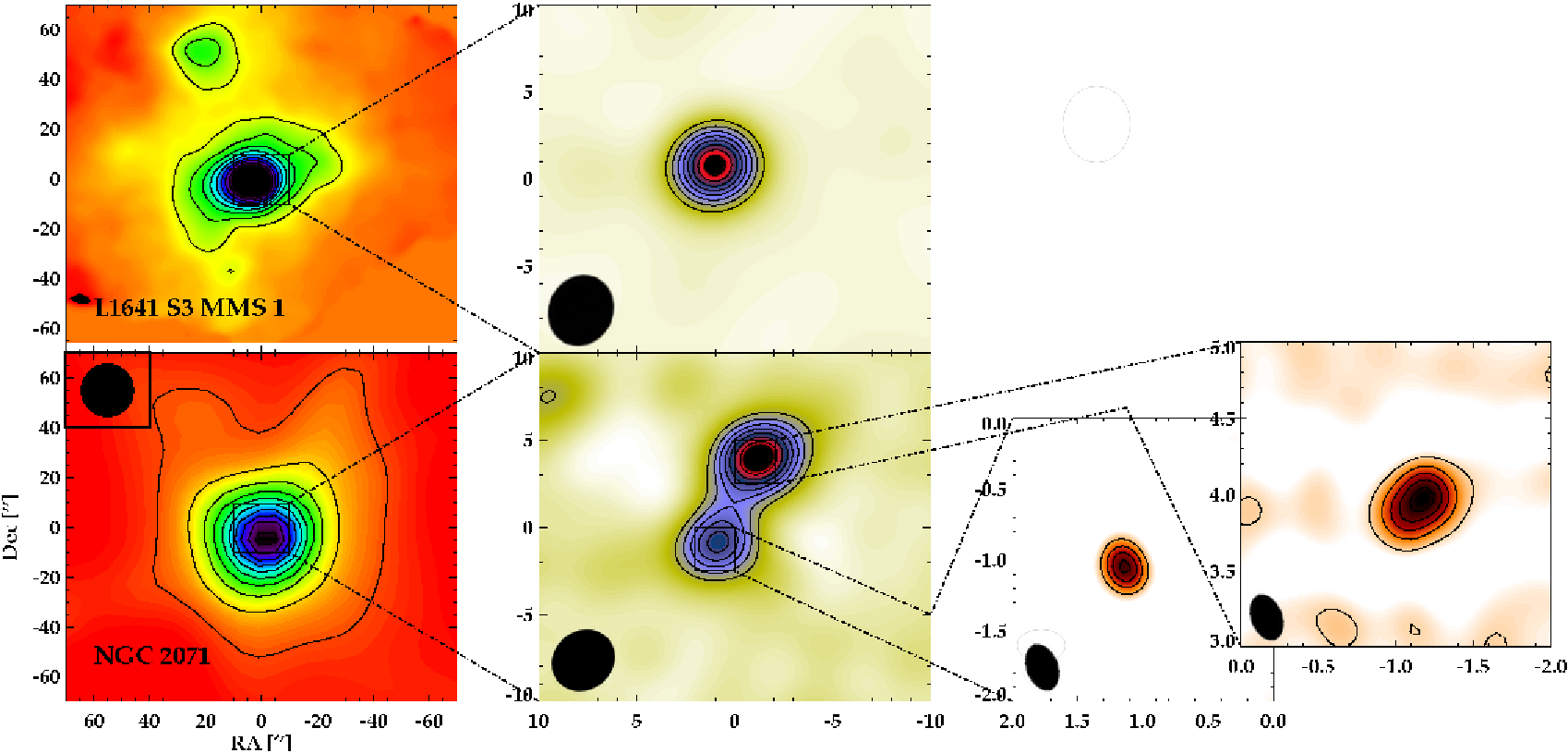}
\end{center}
\caption{Comparison of continuum observations for both Orion sources across multiple resolutions. The first column shows the 850$\mu$m flux from SCUBA with a resolution of $\sim$15$''$. The second column shows the SMA-compact 1300$\mu$m  observations, with a resolution of $\approx$4.5$''$,. The third column shows the very high resolution using the eSMA (NGC 2071 only) at a resolution of 0.3$''$ at a frequency of 350 GHz. Contour lines are shown at 0.1, 0.2... 0.9$\times$ the maximum flux at that resolution, as reported in Table \ref{tab:continuum}. The synthesized beams are shown with black elipses, while the SCUBA beam is shown in the figure of NGC 2071. Extended configuration observations of L1641 S3 MMS 1 can be found in Figure 4.}
\label{fig:1:cont}
\end{figure*}

}

\def\placeFigurezoom{
\begin{figure}
\begin{center}
\includegraphics[width=240pt]{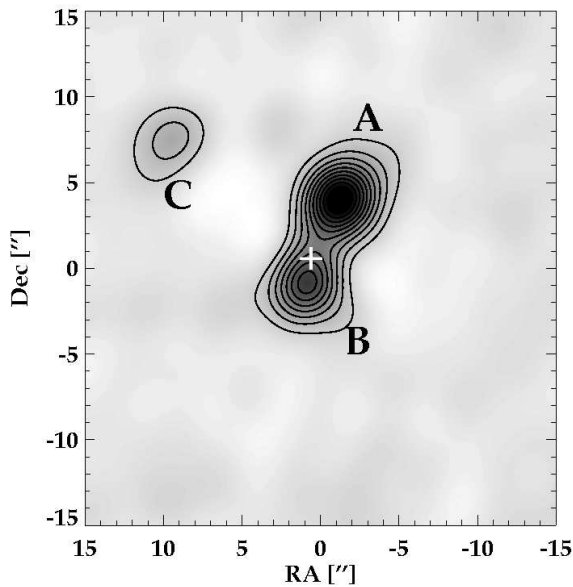}
\end{center}
\caption{Zoomed image of the NGC 2071 protocluster at 230 GHz continuum in the SMA compact configuration. Contour lines are in 3,6,9... $\sigma$ with $\sigma$ = 0.013 Jy/beam. Three cores are identified, labelled A, B, and C and referred to in the text as 2071-A, 2071-B, and 2071-C respectively. The phase center is indicated with a white plus sign.}
\label{fig:2:zoom2071}
\end{figure}
}

\def\placeFigurezoomtwo{
\begin{figure}
\begin{center}
\includegraphics[width=240pt]{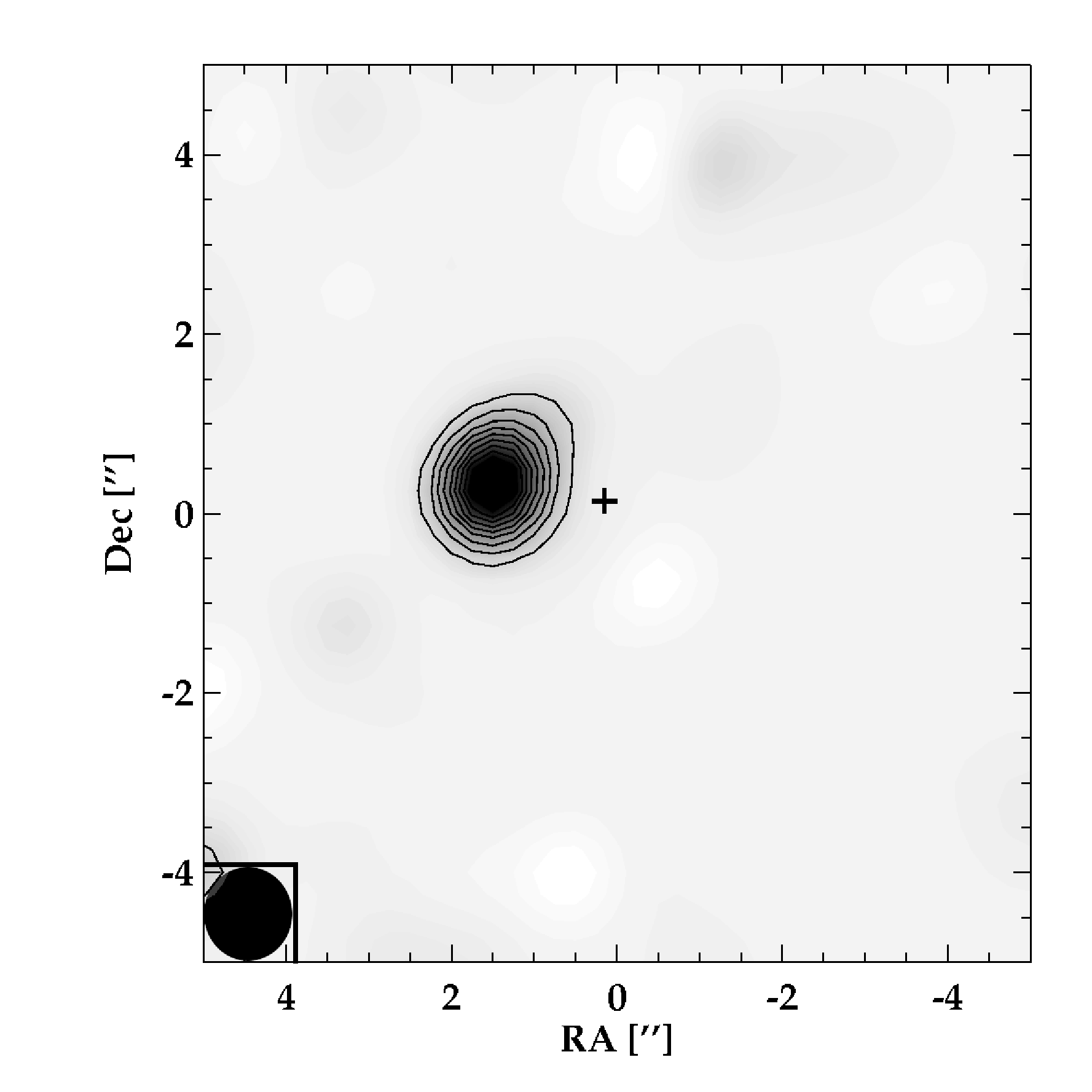}
\end{center}
\caption{Zoomed image of the L1641 protostar at 230 GHz continuum in the SMA extended configuration. A plus sign indicates the phase center. Contour levels are at 10$\%$, 20$\%$, 30$\%$, ...  of the peak flux (0.26 Jy/beam) }
\label{fig:2:zoom1641}
\end{figure}
}

\def\placeFigureuvamp{
\begin{figure*}
\begin{center}
\includegraphics[width=230pt]{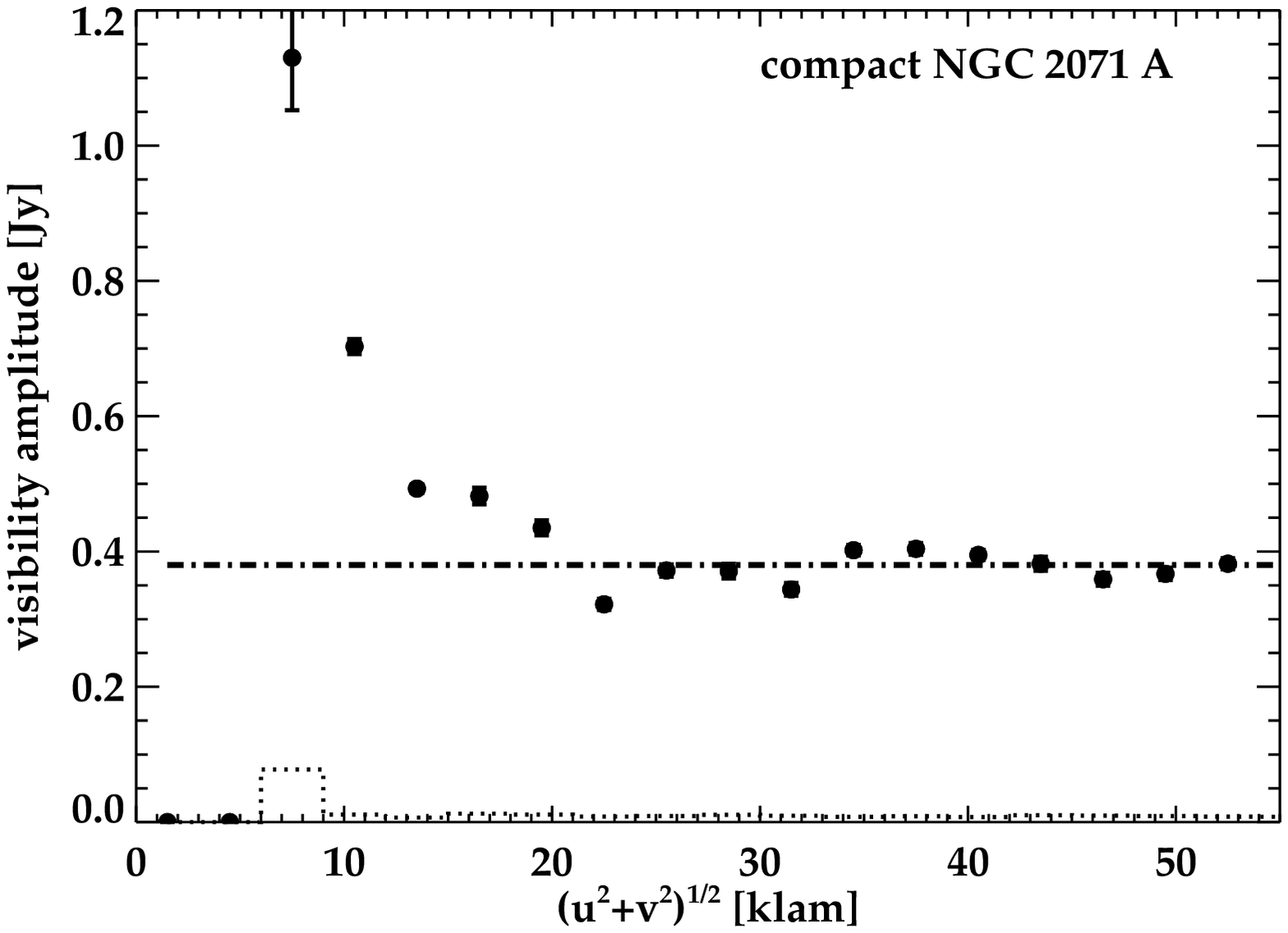}
\includegraphics[width=230pt]{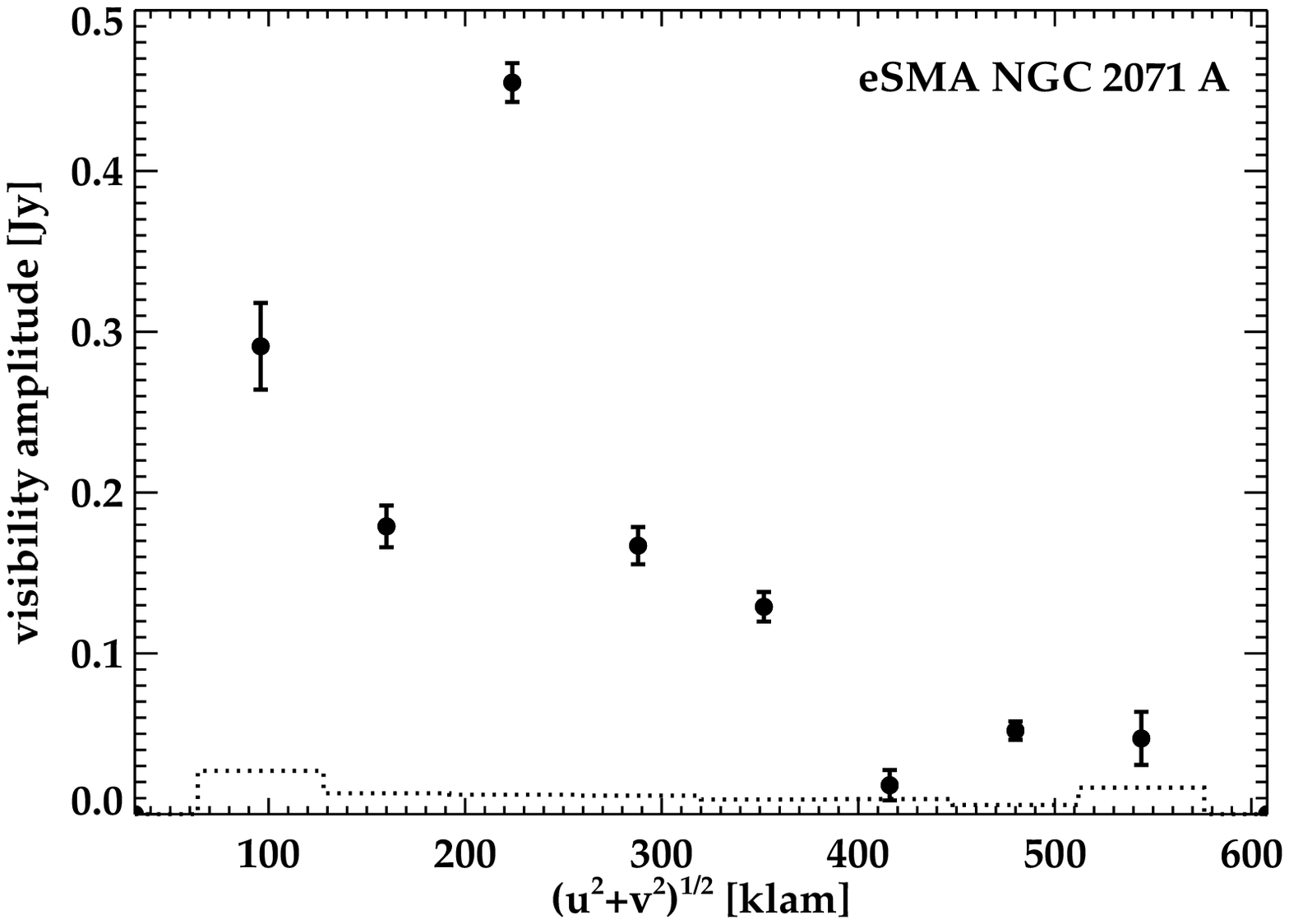}
\includegraphics[width=230pt]{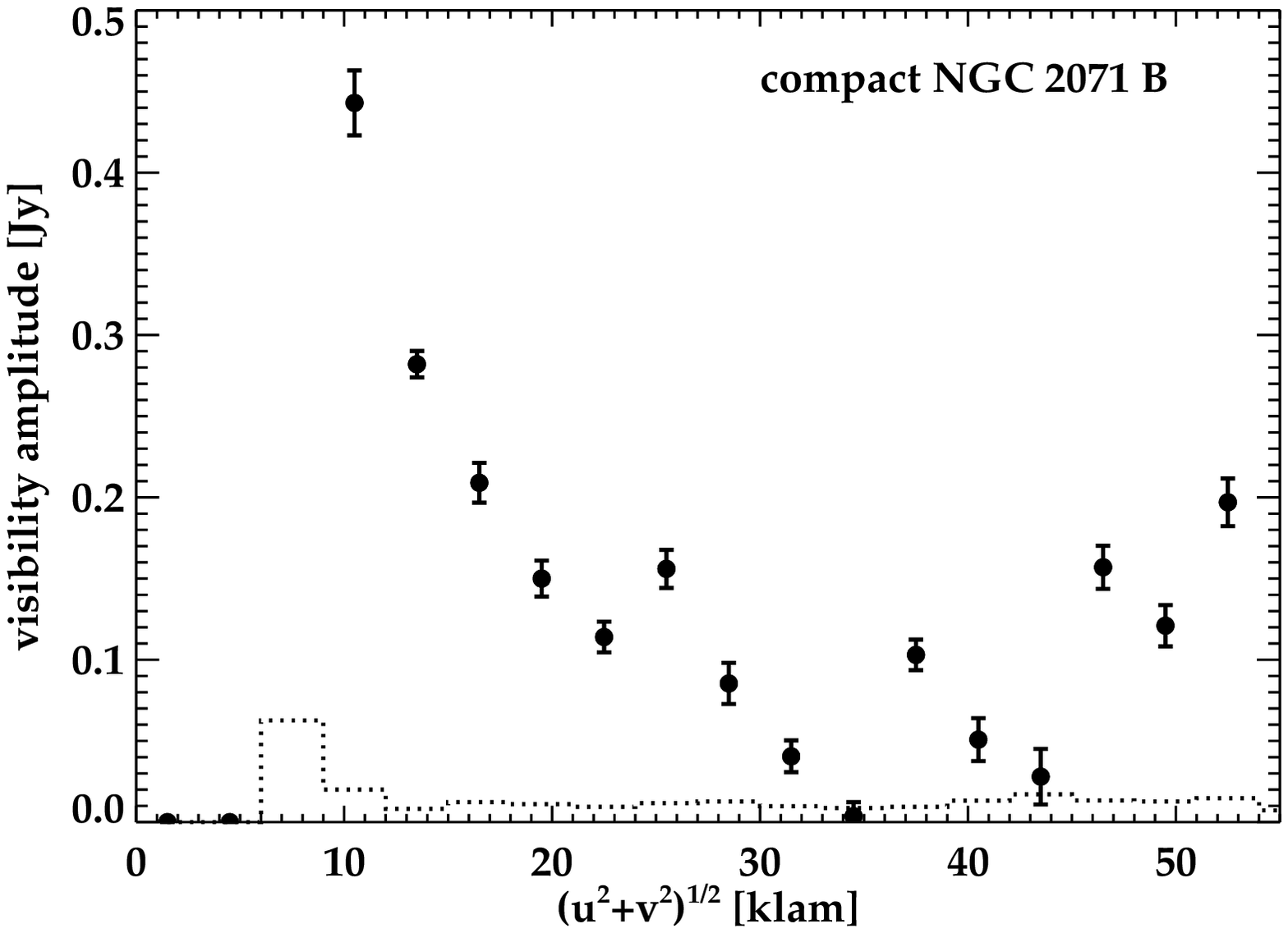}
\includegraphics[width=230pt]{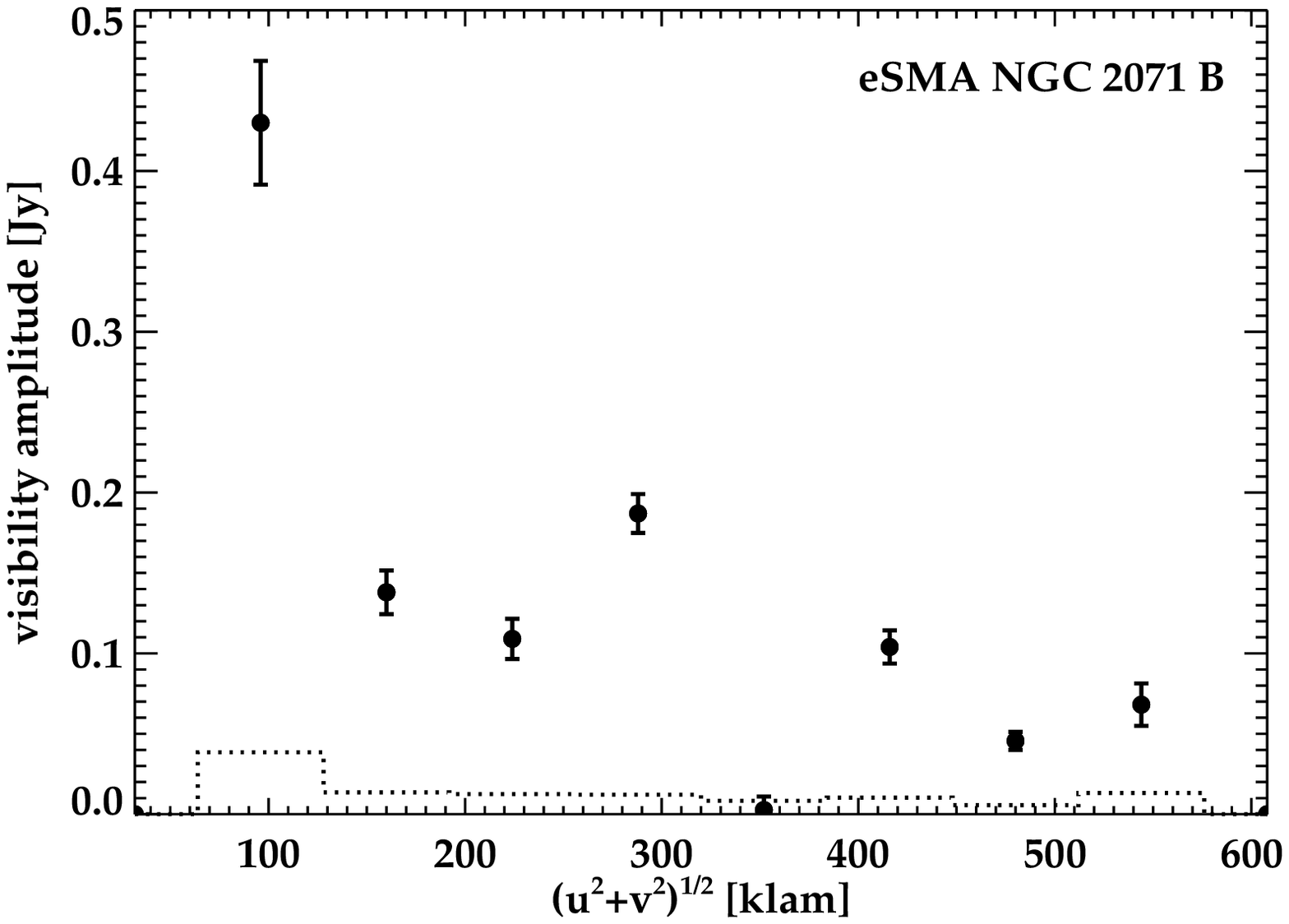}
\end{center}
\caption{Observed continuum visibility amplitudes as a function of the projected baseline length for  2071-A and 2071-B. Error bars are 1 $\sigma$ statistical errors and the dotted histogram indicates the zero-expectation level at 1 $\sigma$. {\it Left column:} The compact configuration data of sources 2071-A ({\it up}) and 2071-B ({\it down}). The data clearly indicate a classical profile as seen in e.g. Fig. 3 of \citet{Jorgensen05} where a resolved envelope transitions into an unresolved disk-like component. For 2071-A in the compact configuration this is indicated by a horizontal line. {\it Richt column:} The eSMA data for the two sources. Here it can be seen that the disks are resolved and the visibilities fall off until  k$\lambda \sim$ 500, or about 80 AU at the distance of Orion.}
\label{Fig:uvamp}
\end{figure*}

}

\def\placeFigureuvampL1641{
\begin{figure}
\begin{center}
\includegraphics[width=230pt]{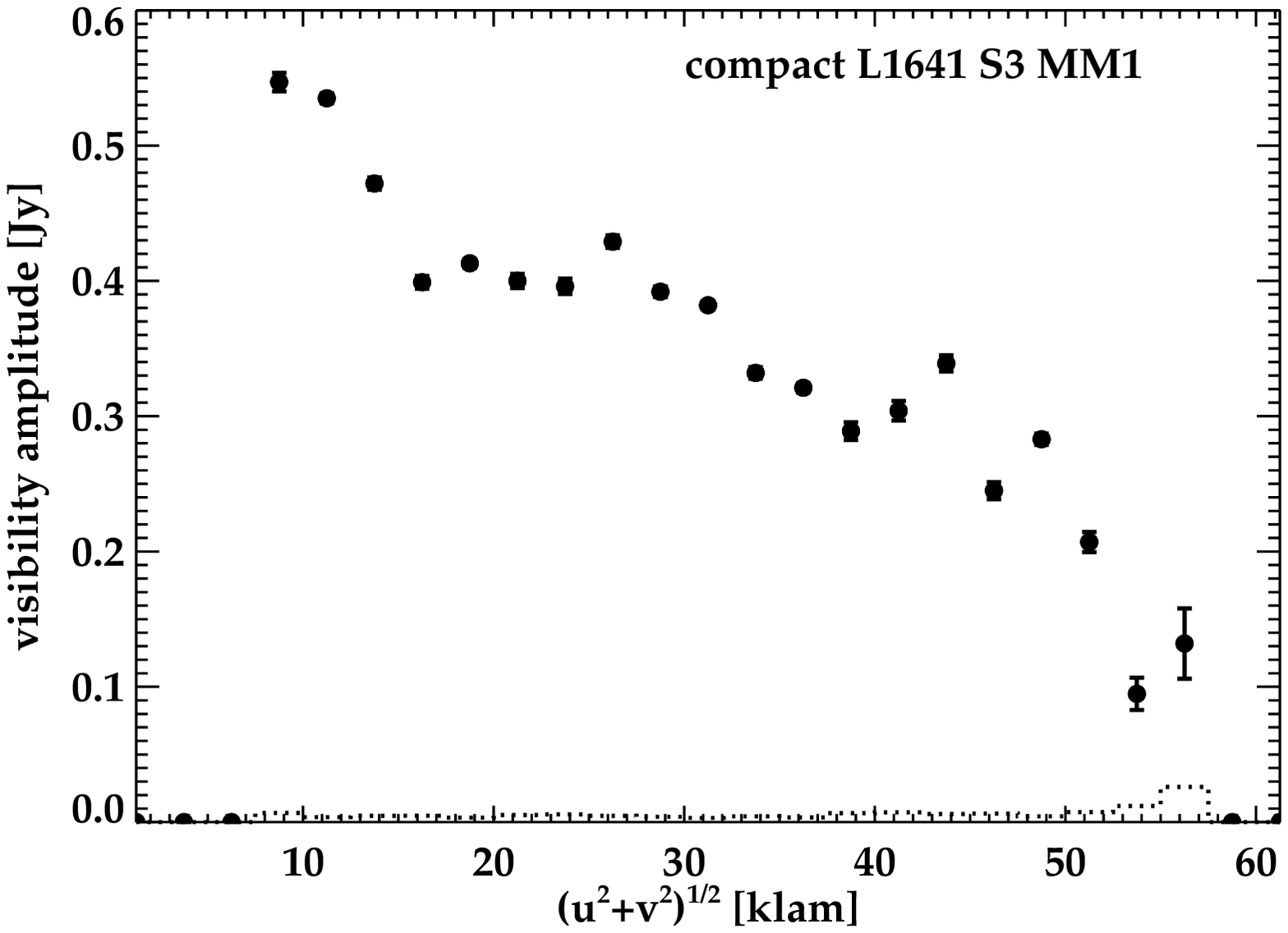}
\end{center}
\caption{Observed continuum visibility amplitudes as a function of the projected baseline length for L1641 S3 MMS 1 in compact configurations. Error bars are 1 $\sigma$ statistical errors and the dotted histogram indicates the zero-expectation level. The extended configuration visibility was found to be unresolved at a level of 0.2 Jy.}
\label{Fig:uvampL1641}
\end{figure}

}

\def\placeFigurefragment{
\begin{figure*}
\begin{center}
\includegraphics[width=500pt]{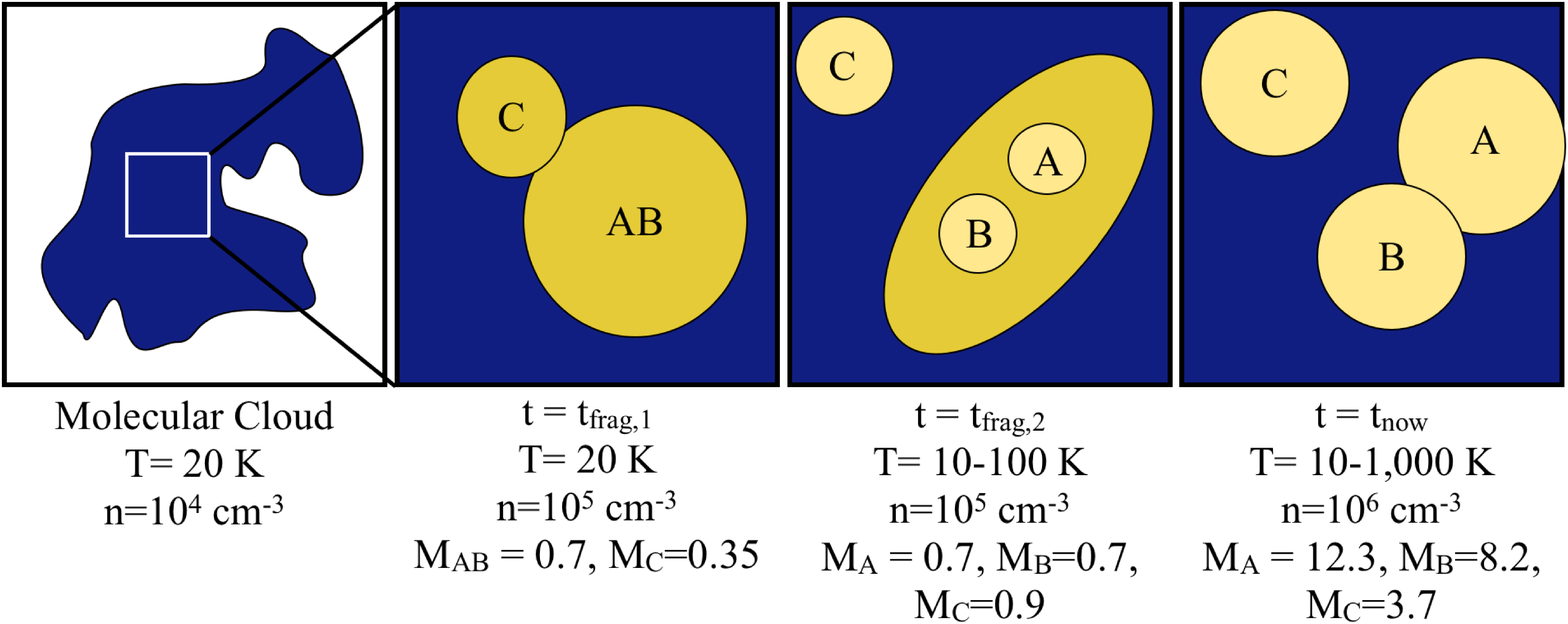}
\end{center}
\caption{Cartoon representation of the fragmentation of NGC 2071. Masses are in units of \Msol.}
\label{fig:frag}
\end{figure*}
}


\def\placeTablesettings{
\begin{table*}[!th]
\begin{center}
\caption{Observational parameters.}
{\footnotesize
\begin{tabular}{l l l l l l l l l l}
\hline \hline
Dates & Config. & Source & Beam size & Bandwidth & Freq. & SB & Band &  $\#$ Chan. \\
 &  & & ($''$x$''$) & GHz & GHz &  & & \\
\hline
Jan 3 2010  & Compact & NGC 2071      & 3.4$"$x2.9$"$& 4 GHz & 230.538 & USB & s13 & 128 \\
Jan 4 2010  & Compact & L1641 S3 MMS1 & 3.5$"$x3.0$"$&4 GHz & 230.538 & USB & s13 & 128  \\ 
Feb 11 2010 & eSMA    & NGC 2071      & 0.37$"$x0.19$"$&2 GHz & 349.415 & USB & s17 & 128 \\
Sept 28 2010 & Extended & L1641 S3 MMS 1 & 0.9$"$x0.87$"$ & 4 GHz & 230.538 & USB & s17 & 128 \\ \hline
Date & Config & Source & Bandpass& Flux Cal.& \multicolumn{3}{c}{Gain Calibrators}  &  \\ \hline
Jan 3 2010 & Compact & NGC 2071 & 3c454.3 & Uranus & \multicolumn{3}{c}{0501-019, 3c120} \\
Jan 4 2010 & Compact & L1641 S3 MMS 1 & 3c273 & Titan & \multicolumn{3}{c}{0607-085,0423-013} \\
Feb 11 2010 & eSMA & NGC 2071 & 3c273 & Vesta, Titan & \multicolumn{3}{c}{Vesta} \\
Sep 28 2010 & Extended & L1641 S3 MMS 1 & 3c454.3 & Callisto/Nept. & \multicolumn{3}{c}{0607-085,0609-157}\\

\hline
\end{tabular}}
\label{tab:setting}
\end{center}
\end{table*}
}

\def\placeTablesources{
\begin{table*}[!th]
\begin{center}
\caption{Source Properties}
\begin{tabular}{l l l l l l l l l }
\hline \hline
Source & R.A. & Dec. & $V_{\rm{LSR}}$ & L$_{\rm{bol}}$$^a$ & Dist.$^a$ \\ 
& hms & dms & km s$^{-1}$ & L$_\odot$ & pc.   \\ \hline
NGC 2071 & 05 47 04.7 & +00 21 44 & 9.6 & 520 & 422 \\
L1641 S3 MM1 & 05 39 56.1 & -07 30 28 & 5.3 & 70 & 465 \\
\hline 

\end{tabular}
\label{tab:sources}
\end{center}
$^a$ Luminosities and distances are adopted from the WISH list \citep{vanDishoeck11} and references therein.
\end{table*}
}

\def\placeTableContinuumresults{
\begin{table}
\begin{center}
\caption{Submillimeter Continuum Measurements}\label{tab:continuum}
\begin{tabular}{l l l l l l}
\hline \hline
Source & Core & Offset & Peak & Total Flux \\
 & & &  Jy/Beam  & Jy \\ \hline
\multicolumn{4}{c}{SCUBA  850 $\mu$m}\\ 
NGC 2071 & - & - & 5.8 +/- 0.8 & 21.9 +/- 4.0\\ 
L 1641 S3 &- & - & 5$^1$ +/- 1.0 & 30.0 +/- 6.0 \\ \hline
\multicolumn{4}{c}{Compact  230 GHz} \\ 
NGC 2071 & A & 4.2$''$ & 0.45 +/- 0.1 & 5.6 +/- 1.1\\
& B  & 1.4$''$ & 0.30 +/- 0.06 & 3.7 +/- 0.7\\ 
 & C  & 12.2$''$ & 0.14 +/- 0.04 & 1.7 +/- 0.4 \\
L1641 S3 MMS 1 &  & 1.4$''$ & 0.42 +/- 0.09 & 5.2 +/- 1.1\\ 
\hline
 \multicolumn{4}{c}{Extended  230 GHz}\\
 L1641 S3 MMS 1 &  & 1.6$''$ & 0.26+/- 0.05  & 1.8 +/- 0.4\\ 
 \hline
\multicolumn{4}{c}{eSMA  349 GHz} \\ 
NGC 2071 &  A & 4.2$''$ & 0.13 +/- 0.04 & 0.35 +/- 0.1\\
 & B& 1.4$''$ &0.12 +/- 0.04  & 0.3 +/- 0.1\\
\hline
\end{tabular}\\
\end{center}
$^1$ The SCUBA flux L1641 S3 MMS 1 was estimated from the raw data. See text. Earlier papers \citep{Snell86,Trinidad09,Carrasco12} identify 2071-A with IRS-3, 2071-B with IRS-1 and 2071-C with IRS 2. 

\end{table}
}

\def\placeTableSpitzer{
\begin{table*}
\begin{center}
\caption{Spitzer and 2MASS Photometry}\label{Spitz:tab}
\begin{tabular}{lllllllll }
\hline \hline
Source & J & H & K & IRAC 1 & IRAC 2 & IRAC 3 & IRAC 4 \\  
& Jy & Jy & Jy & Jy & Jy & Jy & Jy \\
\hline
2071-A & 1.6e-4 & 7.9e-4 & 4.6e-3 & 7.0e-2 & 4.9e-2 & 2.2e-1 & 0.9\\
2071-B & 3.0e-4   & 2.2e-3 & 2.1e-2 & 4.4e-2 & 3.7e-2 & 1.7e-1 & 0.7  \\
2071-C & 2.3e-3 & 6.0e-3 & 4.5e-2 & 8.6e-2 & 6.0e-2 & 2.7e-1 & 1.2 \\
L1641 S3 MMS 1 & - & - & - & 0.69 & 0.82 & 2.3 & 17.2 \\ \hline
\end{tabular}
\end{center}

\end{table*}
}

\def\placeTablemassresults{
\begin{table}
\begin{center}
\caption{Mass estimates}\label{tab:masses}
\begin{tabular}{l l l l l l}
\hline \hline
Source &  & 15$''$ & 4$''$$^a$ & 0.9$''$$^a$ & 0.25 $''$\\
 & & \multicolumn{3}{c}{Mass [M$_{\rm{\odot}}$]} \\ \hline
NGC 2071 & Core$^b$ & 21.7 & - & - \\
& A &- &12.3 & -&0.35 \\
& B &- & 8.2 &-&0.29 \\
& C & -& 3.7 & -&$<$0.2 \\
\hline
L1641 S3 MMS1 &- & 20.9 & 11.46 & 3.9 \\ 
\hline
\end{tabular}\\
\end{center}
$^a$ Observations taken at 230 GHz with dust opacity $\kappa$=0.009 cm$^2$ gm$^{-1}$. \\
$^b$ Mass measurement derived for the single protocluster core.\\

\end{table}
}

\def\placeTableSED{
\begin{table*}
\begin{center}
\caption{Results from Robitaille SED fit$^a$}\label{tab:SED}
\begin{tabular}{l l l l l l l l l l}
\hline \hline
Source & \Lbol & Env. Mass$^b$ & Stellar Mass & (M$_{env}$/M$_{star})^b$ \\
& \Lsol & \Msol  & \Msol \\ \hline
2071-A         & $<$27  &  8.2      & 0.9     & 30 \\
2071-B         & 10  & 14.2      & 0.5    & 20\\
2071-C         & 3.4 & $\sim$3.5 & 0.5    & 6.8\\
L1641 S3 MMS 1 & 250 & $>$9.6    & $>$3.5 & 2.7\\
\hline
\end{tabular}
\end{center}
$^a$ Results are averaged over the 10 best-fitting models. The spread in these values is about 50$\%$ from the given value. \Av values range from 5 - 35 and are the main cause for the large uncertainty.\\
$^b$ Lower limit estimates.

\end{table*}
}

\def\placeTablefragment{
\begin{table}
\begin{center}
\caption{Fragmentation parameters and results.}
\begin{tabular}{l l l l l l} 
\hline \hline
Core Combination & Distance & \multicolumn{2}{c}{Scenario I} & Scenario II  \\
& & $\Lambda$ & $M$ \\
 & AU & cm$^{-3}$ & cm${-3}$ &  \\
\hline
AB & 2430 & 7.4e5 & 1.3e2 \\
AC & 5131 & 1.7e5 & 1.6e3 \\
BC & 5420 & 1.5e5 & 1.6e3 \\ \hline
\end{tabular}
\end{center}
\label{tab:frag}
$^a$ an isothermality of 20 K is assumed. As such T$_{\rm{frag}}$ is not a calculated paremeter.
\end{table}
}


\section{Introduction}
    Star formation research is a cornerstone of current-day galactic astronomy. A solid understanding of star formation allows us to analyze astronomical structures over a wide range of physical scales,  from  disk and (giant) planet formation around sunlike stars to the physical structure of the giant molecular clouds which form the building blocks of galaxies and which play an important role in galaxy evolution. Star formation studies are necessary in order to define the initial conditions of most objects in our universe. Typically, galactic studies of star formation concentrate on either isolated low-mass (LM) \citep{Lada84,Lada87,Andre93,Shirley00,Jorgensen04,Bottinelli04,Evans09} or high-mass (HM) star-forming regions, which are, with few exceptions, seen almost exclusively in clusters \citep[e.g.][]{Beuther05,Beuther07, Ragan09, Smith09,Keto10}. Both ends of the mass spectrum provide unique perspectives on star formation. Many isolated LM star-forming regions have the advantage of being nearby, allowing them to be observed at high physical resolution, and are far less affected by strong radiation fields created by neighbouring (proto)stars. These LM stars are often assumed to form within individual collapsing envelopes \citep{Shu77}. Most field stars in our galaxy \citep[see e.g][]{Bressert10}, however, form in clusters where interactions from nearby more massive stars, stronger radiation fields, and multiplicity due to fragmentation in the parental cloud complicate the star formation process 
\citep{Adams01,Adams06,Duchene07}.  Although physical models have existed for a decade \citep[e.g.][ and references therein]{McKee99,Klessen01,Krumholz05, Bate05,Vasquez09}, only a handful of recent studies have tried to observationally determine cluster properties \citep{Smith09,Longmore11}.
In broad terms, the evolution of isolated LM protostars is reasonably well understood \citep{Lada87,Andre93, Robitaille06, Crapsi08,vanKempen09}. Even with the recent discovery of the VeLLO class, significant progress has been made to include these types of protostars in the general evolutionary picture \citep{Dunham10,Vorobyov10,Vorobyov11}.  Recently an evolutionary picture has been identified for very massive protostars \citep{Fontani09,Keto10}. No systematic effort, however, has been undertaken to observationally confirm the theories of clustered star formation that have been put forward by numerical modelling \citep[e.g.][]{Klessen01,Bate05,Bate09}.

 Recent work from e.g. \citet{Keto10} and 
\citet{Johnston11} show that there may be significant similarities in the formation mechanism of individual 
HM and LM protostars. As an example, for star formation at masses above M $>$ 8 M$_\odot$ there has been concern that radiation pressure might choke off the accumulation of mass from a surrounding envelope but recent results suggest that this is mitigated by the physical structure and geometry of the region, allowing O stars to form through core collapse \citep[e.g.][]{Krumholz06,Kuiper10}. The main observed differences in  the star formation process all have an origin in the environment and the energies involved in accretion and outflows. These can be enumerated as: (i) the clustering of LM protostars around HM protostars and thus the influence of the environmental radiation field on LM protostellar formation \citep{Krumholz10} (ii) the fragmentation of natal envelopes before nuclear fusion begins in the heaviest members \citep[e.g.][]{Bontemps10} (iii) the strength of the internal  radiation field produced by different stellar surface temperatures and its influence on the surroundings, and (iv) the feedback from powerful shocks created by outflow interactions with the parental cloud, both inducing and dampening star formation \citep[e.g][]{Arce06}.

Intermediate mass (IM) protostars (defined observationally through their bolometric luminosity: \Lbol $>$ 50 and $<$ 2,000) have not been studied in detail, but may reveal crucial information on the differences between the modes of star formation.
Although a few individual sources have been observed and analyzed in great detail [e.g. NGC 7129 IRS 2,
\citet{Fuente05a,Fuente07}, IRAS 20050+2720, \citet{Beltran08}], it is uncertain if these are 
either typical of their evolutionary stage/age or  typical of clouds forming more massive stars than those in nearby star-forming regions, e.g. Taurus or Ophiuchus. Despite these limitations, IM protostars make excellent test-cases for star formation theories that aim to include the full range of stellar masses \citep[e.g.][]{Fontani09,Kama10,Palau10}.     
Being more
luminous than LM protostars, these objects provide for larger warm zones within the enshrouding envelopes. These warm zones around forming stars were once thought to be nearly spherical in nature (the hot core), however, observations and models of key molecular lines now indicate that at least for LM protostars the structure of the inner envelope and outflow cavity walls plays an important role  \citep{vanKempen09a,Bruderer09,Visser12}. 
 Recent observations \citep{vanKempen09a,vanKempen10a} show that this shell structure provides an accurate accounting of the heating and cooling balance and thus the evolution of low-mass protostars. Due to the large distance ($\sim$ 2 kpc) to most HM protostars, resolving the structure in such warm regions has proven difficult. 
IM protostars are natural laboratories  that can probe possible differences between LM and HM protostars as they can be found at much smaller distances  (0.5-1 kpc) than HM protostars and yet they produce significantly more UV photons than LM protostars. 

A second important characteristic of star formation theories can also be studied with IM protostars due to their proximity. Small mini-clusters of LM protostars cannot be distinguished from single HM protostars at large distances ($>$ 2 kpc) with even the highest resolutions of the current generation of sub-mm interferometers. 
There are few observational papers investigating clustering and attempting to quantify the fragmentation of HM cores into multiple members e.g. \citet{Beuther04,Brogan09,Longmore11}. IM protostars in Orion are near enough to allow facilities such as the SMA, the IRAM PdB interferometer, CARMA and in the future ALMA, to individually distinguish cluster members and thus constitute an intriguing sample.

This paper presents an analysis of the small-scale physical structure of two protostars in Orion, the nearest cloud producing protostars more luminous than 50 L$_\odot$.  The two observed sources, NGC 2071 and L1641 S3 MMS 1 \citep{Seth02,Stojimirovic08,Skinner09} were selected specifically because (i) they are more massive than typical low-mass protostars (e.g. the PROSAC survey in \citet{Jorgensen05,Jorgensen07} and protostellar surveys in Ophiuchus \citep{Johnstone00, vanKempen09}), (ii)  they are located in the Orion cloud, where the ISRF is significantly larger than in other nearby regions such as Taurus and Ophiuchus \citep{Jorgensen06}, and (iii) they are still near enough ($\sim$ 450 pc) to allow the  reasonable identification of circumstellar disks \citep[e.g.][]{McCaughrean96,Seth02,Menten07} and protostellar envelopes. These two sources 
are not located in the most crowded region near Orion-KL, making it easier to separate the protostars from their surroundings. Further, these sources are thought to be representative of intermediate mass star formation in general and therefore excellent test cases to probe star formation in the mass range between the low and high-mass ends.  In this paper, we present the continuum observations and discuss the physical structure of these two sources. The format is as follows: $\S$ 2 discusses the observations, the results are presented in $\S$ 3, $\S$ 4 analyses the data, and the structure of IM protostars is discussed in $\S$ 5. Finally, conclusions are presented in $\S$ 6.

\placeTablesources
\placeTablesettings

\section{Observations}
The two Orion sources investigated in this paper, NGC 2071 and L1641 S3 MMS1, were selected based on the proposed source list of the Herschel Guaranteed Time key program WISH (Water in Star-forming region with Herschel,  PI: E.F. van Dishoeck) subprogram on intermediate mass protostars \citep[see e.g.][]{Fich10,Johnstone10,vanDishoeck11}. At a distance of $d$ estimated around $\sim$ 450 pc \citep{Johnstone01}, Orion is the only active star-forming region producing massive stars within 500 pc \citep{Sadavoy10,Buckle10} 
and is thought to be the only cloud at such distances which is producing IM protostars. Note that the distances to individual regions of Orion might vary. The OMC might be as close as 414 pc \citep{Menten07}. As such the error of this distance is likely as large as 40 pc. NGC 2071 is a much-studied area \citep{Snell86, Torrelles98} and was very recently reported on by \citet{Carrasco12} using cm and 3 mm wavelength observations at very high resolution.
The known properties of the two selected sources can be found in Table \ref{tab:sources}. 
The two Orion sources were observed with the Submillimeter Array\footnote{The Submillimeter Array
 is a joint project between the Smithsonian Astrophysical Observatory and the Academia Sinica Institute of Astronomy and Astrophysics and is funded by the Smithsonian Institution and the Academia Sinica.} over three nights in January/February 2010 using the compact and eSMA configurations. Further observations of L1641 S3 MMS1 were obtained in September 2010 using the extended configuration of the array.
These SMA observations are complemented by continuum observations from 2MASS, Spitzer and the JCMT.

\placeFigureSCUBA

\subsection{SMA}
The Orion sources were observed in two campaigns on the SMA covering three configurations: compact, extended, and eSMA. Table \ref{tab:setting} describes the settings that were used during each of the observations, including beam sizes, bandwidths, correlator configurations as well as the bandpass, amplitude and gain calibrators used for the different dates. \\
Initial observations were made in the compact configuration on 3$^{\rm{rd}}$ and 4$^{\rm{th}}$ January 2010. On February 11$^{\rm{th}}$, observations  were made in the eSMA configuration. In the eSMA configuration\footnote{The eSMA (extended SMA) is a collaboration of the SMA, JCMT and CSO, to join the three facilities into a single long baseline sub-mm interferometer. For more information on the eSMA, see \citet{Bottinelli08} and \citet{Shinnago09}} the  very extended configuration of the SMA is combined with the James Clerk Maxwell Telescope (JCMT)\footnote{The JCMT is operated by The Joint Astronomy Center on behalf of the Science and Technology Facilities Council of the United Kingdom, the Netherlands Organization for Scientific Research, and the National Research Council of Canada.} and the Caltech Submillimeter Observatory (CSO).  Additionally, L1641 S3 MMS1 was observed in a filler on September 28th 2010 using 6 dishes at 230.358 GHz in extended configuration.

For the compact and extended configurations, instead of the normal 2 GHz correlator bandwidth, the enhanced double-bandwidth mode was employed. In this mode, 4 GHz is obtained in both the lower and upper sideband for a total of 8 GHz bandwidth. The main molecular lines targeted were $^{12}$CO 2-1, in the upper sideband (USB), and the two well-studied isotopologues $^{13}$CO and C$^{18}$O, in the lower sideband (LSB). An analysis of these CO lines will be presented in a future paper.

For NGC 2071, the quasars 0501-019 and 3C120 were used as gain calibrators and Uranus as a flux calibrator. For L1641 S3 MMS 1, the quasars 0607-085 and 0423-013 were used as gain calibrators, and Titan as a flux calibrator. During the eSMA observations, the correlator setup was limited to the 2 GHz bandwidth mode due to the increased number of antennas. The limited observing time of 1.5 hours  was nevertheless sufficient to cover a significant part of the uv-plane. 3c273 was used as bandpass and Titan and Vesta as flux calibrators. Due to its close proximity, Vesta was also used as a gain calibrator. The correlator was fixed at 349 GHz, due to track-sharing with observations of Titan.

The data reduction was performed using a combination of the following software tools: the MIR package for IDL, MIRIAD, and CLASS in GILDAS\footnote{GILDAS is a software package developed by IRAM to reduce and analyze astronomical data. http://www.iram.fr/IRAMFR/GILDAS}. 
For all images, cleaning was done to 3$\sigma$ using the Clark method and a natural weighting was applied  to obtain the best images for our goals. Although uniform weighting can provide better image fidelity, the price of the lower signal to noise and thus the detection of multiple components as well as their relative strengths, was considered not to be in line with the goals.

\subsection{Complementary data}
Several other astronomical data-sets are utilized in this paper as an aid to analyzing the source properties. SCUBA 850 $\mu$m continuum flux archive data \citep{diFrancesco08} is available for NGC 2071.  Unfortunately, L1641 S3 MMS 1 was not included in the SCUBA archive due to its uncertain calibration. It was, however, observed by SCUBA and measurements were retrieved. \citet{Zavagno97} observed L1641 S3 MMS 1 using the predecessor of SCUBA, the UKT14 common user bolometer instrument, and reported a calibrated flux of 5.14 \, Jy at 800 $\mu$m. Comparing the UKT14 results, corrected for the observed source size, and the uncalibrated SCUBA observations allow us to adopt a total flux of 5$+/-$1 Jy at 850$\mu$m for L1641 S3 MMS1 and to calibrate the SCUBA map.
Spitzer  IRAC and MIPS 24 photometry were obtained for both sources from the Spitzer archive (Megeath, priv. comm). The data was taken within the scope of the Orion Spitzer program [PI: Tom Megeath, see e.g. \citet{Gutermuth09} and Megeath, T. et al. (in prep) for more information].
Near-IR photometric data points for the SED analysis were obtained from the 2MASS archive. The continuum measurements are provided in Tables \ref{tab:continuum} and \ref{Spitz:tab}.

\placeTableContinuumresults

\placeTableSpitzer

\section{Continuum Results}
For each of the Orion sources Figure \ref{fig:1:cont} shows the submillimeter continuum observations at different spatial resolutions, ranging from 15$''$ to 0.2$''$, using a combination of the 850$\mu$m and 1.3 mm data from the SMA, eSMA and JCMT/SCUBA. The measured fluxes at each of these scales is tabulated in Table \ref{tab:continuum}. In the figure, images are provided at three different length scales (120$''$, 20$''$ and 2$''$) and each image is normalized to the peak flux of the source in order to accurately compare the source structure.  It is clear that the large-scale environments of L1641 and NGC 2071 look very similar at scales of $\sim$30$''$ (Fig. \ref{fig:1:cont}) and the peak fluxes of each core are remarkably close (5 versus 5.2 Jy), within the calibration error of SCUBA (20$\%$).

\placeFigurezoom

\placeFigureuvamp

\subsection{NGC 2071}

Multiplicity is observed with the NGC 2071 core when observed with higher spatial resolution using
 the SMA-compact observations, as shown in the 2nd column of Figure \ref{fig:1:cont}. The source shows two  fragmented peaks, 2071-A and 2071-B, with a third peak, 2071-C, to the north-east (see Fig. \ref{fig:2:zoom2071}), clearly detected (S/N$>$30) in the map which has a 3-$\sigma$ rms noise of 4 mJy/beam. At the highest resolutions available (eSMA at $\sim$840 $\mu$m) individual disks in the NGC 2071 sources are discerned within the 2071-A and 2071-B cores. This eSMA map has a 1-$\sigma$ noise level of 7 mJy/beam. The error in the peak fluxes are dominated by the errors in flux calibrations.These three peaks can be identified with older identifications \citep{Snell86,Torrelles98,Carrasco12}: 2071-A is identified as IRS 3, 2071-B as IRS 1 and 2071-C as IRS 2. The source shown in \citet{Carrasco12} between A and B, first identified as VLA 1 by \citet{Trinidad09} is unresolved in the SMA map, although a non-circular elongation of sources A and B does point towards this position. In the eSMA map, a 3.4 $\sigma$ unresolved signal is seen at the position of VLA 1. 

Figure \ref{Fig:uvamp} shows,  for the cores 2071-A and 2071-B, the observed visibilities from the SMA and eSMA as a function of projected baseline calculated from the flux peaks by using the UVAMP MIRIAD task. To derive these visibilities a gaussian model of the second source was subtracted in the image plane  before calculating the visibilities. The profile for source A in the compact configuration shows a resolved envelope, up to 20 k$\lambda$, combined with an unresolved central source  of 0.38 Jy (similar to Fig. 3 of \citet{Jorgensen05}). Although source B is not as bright as source A, a similar fit can be reached in which an envelope is resolved up to $\sim$30 k$\lambda$,  with an unresolved central component of 0.15 Jy. The eSMA observations confirm that these unresolved components are disks, as they are resolved on scales of a few hundred k$\lambda$  corresponding to spatial scales of $\sim$200 AU at the distance of Orion.  The disks are both elongated, but suggest different major axis directions. Central unresolved components within both 2071-A and 2071-B are constrained to $<$ 0.1 Jy. 

If we extrapolate the 230 GHz fluxes to 345 GHz using a simple blackbody model with a $\beta$ of 2, the total emission picked up in the compact configuration SMA in A, B and C would be ~19 Jy. With the observed flux of 21.6 Jy in SCUBA, we retrieve about 90$\%$ of the total flux. Of course, this is highly dependent on the uncertainty of the absolute flux calibration.

No disk-like component was detected down to a level of 10 mJy/beam in the eSMA observations toward 2071-C. As this position was relatively far off the phase center (20 percent of the SMA antenna's primary beam) and in fact outside the primary beam of the JCMT and CSO, the presence of a disk  cannot be ruled out. Note that \citet{Carrasco12} resolve 2071-C as a close binary with a separation of about 500 AU. This would have been detectable with our resolution, but from their Fig 2 this would not be detectable with a 3 $\sigma$ confidence with the sensitivity without the CSO and JCMT. We present a more detailed analysis in $\S$ 4.3.

\subsubsection{Spitzer maps}
\begin{figure*}
\begin{center}
\includegraphics[width=400pt]{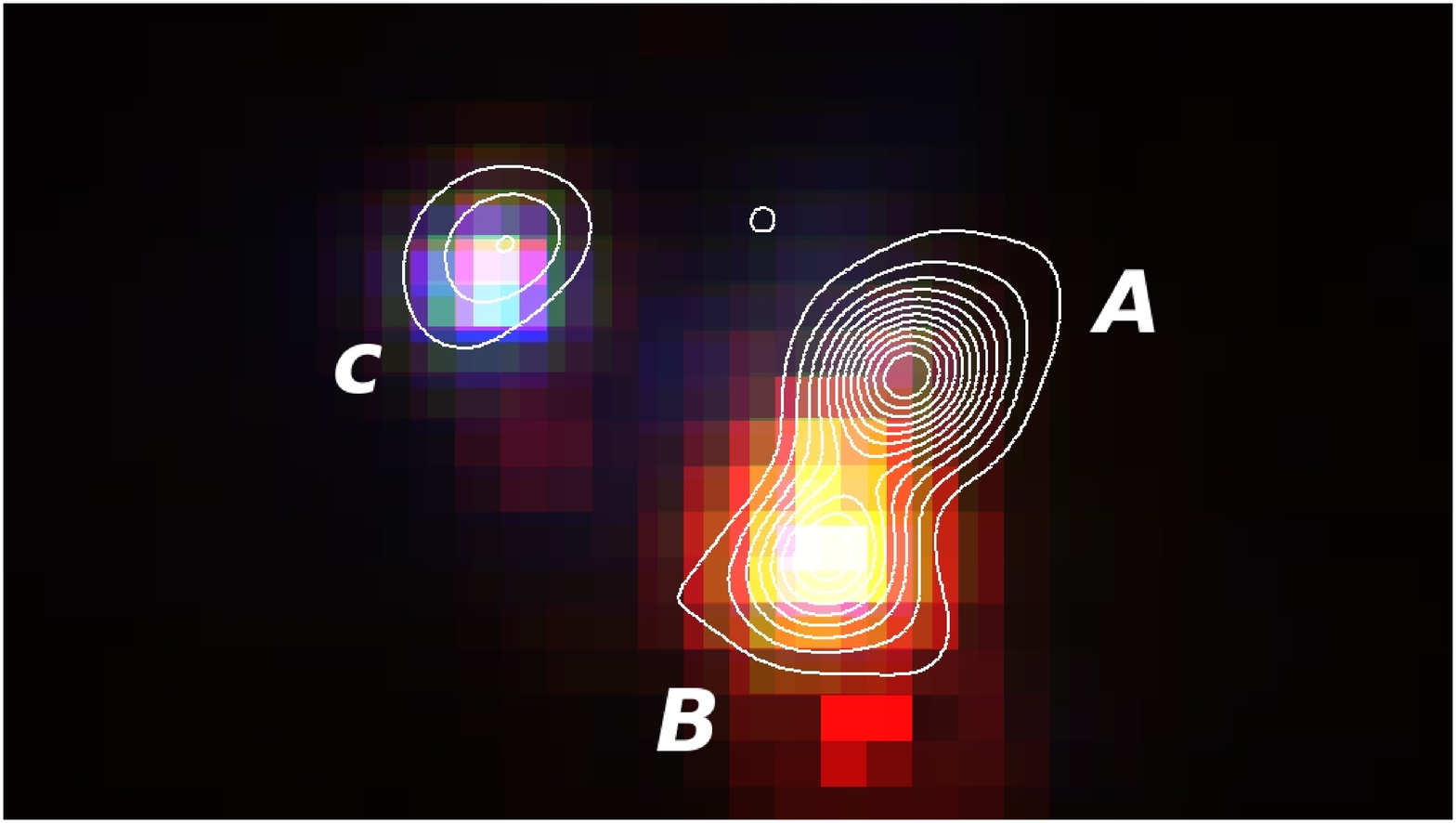}
\end{center}
\caption{SMA continuum observations at 230 GHz at 3,6,9... $\sigma$ plotted in contours of the compact configuration in Fig. \ref{fig:1:cont} and \ref{fig:2:zoom2071} over a Spitzer IRAC 3.6, 4.5 and 8 $\mu$m false color image of NGC 2071.}
\label{Spitz:2071}
\end{figure*}

Spitzer imaging from \citet{Skinner09} can provide important constraints on the individual sources. Over-plotted SMA observations  on the Spitzer-IRAC image are shown in  Figure \ref{Spitz:2071}.

Comparing with the Spitzer data in Table \ref{Spitz:tab} reveals that 2071-B and 2071-C are clearly detected in all bands, although 2071-B is dominated by emission at redder wavelengths. Source 2071-A, which is the brightest in the SMA images, is not clearly detected in any band, with only weak emission at 8 $\mu$m and 24 $\mu$m. 
At the position of 2071-C the source is dominated by the 3.6 $\mu$m emission.

\subsubsection{eSMA and VLA-1}

\begin{figure}
\begin{center}
\includegraphics[width=250pt]{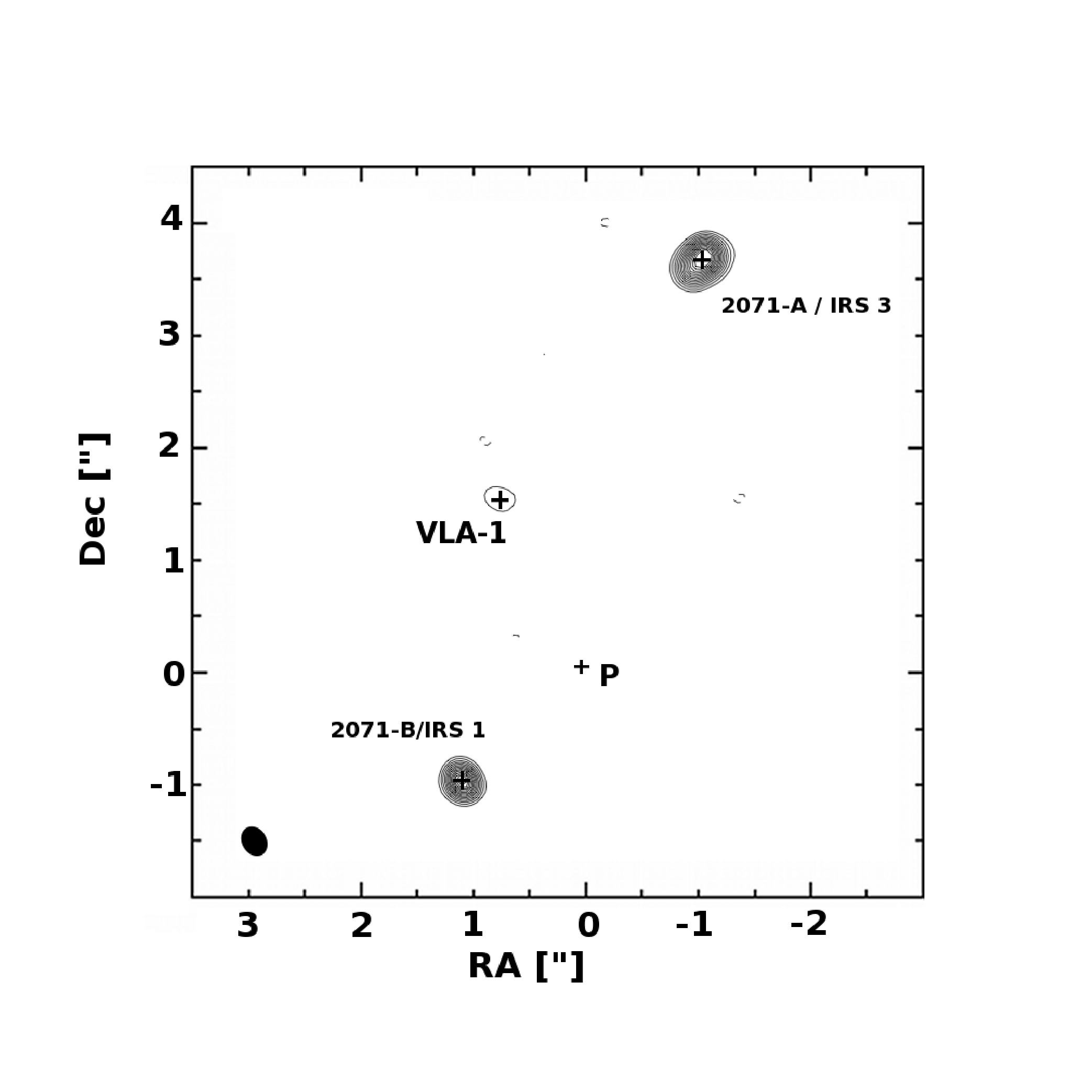}
\end{center}
\caption{The eSMA continuum observations at 349 GHz at a resolution of $\sim$0.25$"$ of the area around 2071-A and 2071-B, identifying these cores, as well as showing the locations of the peak fluxes at 3 mm as shown in \citet{Carrasco12}. In addition the Phase center is shown by a plus sign. Contours are in levels of 3-$\sigma$, with $\sigma$ equalling 7 mJy/beam. The synthesized beam is shown with a black ellipse.}
\label{Fig:eSMA}
\end{figure}

The recent paper by \citet{Carrasco12} identified for the first time the radio source VLA-1 \citep{Trinidad09} at mm wavelengths, with a flux density of 9 mJy. At this position, a signal of 27 mJy, equalling a 3.85$\sigma$ detection, can be seen, likely associated with VLA-1. Fig. \ref{Fig:eSMA} shows the eSMA image of the area around 2071-A, 2071-B and this potential new source.
From the SED of Fig 2 in \citet{Carrasco12} a predicted flux of $\sim$30 mJy can be derived at the eSMA wavelength. The detection at submillimeter would confirm the theorized existence of another protostar in the NGC 2071 cluster. None of our other observations, including Spitzer, has the resolving power to distinguish any emission from VLA-1 from either 2071-A or B. X-ray emission from \citet{Skinner09} has also been associated with this source. We therefore tentatively support the conclusions from \citet{Carrasco12} that VLA-1 is a very young, even more embedded protostar. High-resolution, mid-IR observation are needed to distinguish any IR emission from the region to confirm this however. For the remainder of this paper, we will not discuss this source in any detail due to the limitations of the eSMA detection and observations at other wavelengths. Any other potential clumps similar to VLA-1 are not seen down to our detection limit. 
\placeFigurezoomtwo
\placeFigureuvampL1641

\subsection{L1641 S3 MMS 1}
In contrast to NGC 2071, Figure \ref{fig:1:cont} reveals that L1641 S3 MMS 1 is not fragmenting into multiple sources  at small scales. Only a single continuum source is seen at a size scale of 4.5$"$.  The 1.3 mm continuum emission from the SMA compact configuration shows an almost perfectly circularly symmetric source. In extended configuration, Figure \ref{fig:2:zoom1641}, the source looks unresolved , with a peak flux of 0.26 Jy/beam, roughly half of the peak seen in compact configuration of 0.45 Jy/beam. \textbf{The central unresolved component is likely associated with  disk with radius $<$300 AU }while the more than 50$\%$ envelope is significantly filtered out by the extended configuration. It is also clear that the original coordinates of L1641 S3 MMS 1 are off by almost 2 arcseconds. The new updated position is 05$^h$39$^m$56.1$^s$, -07$^d$30$^m$28$^s$. The original coordinates were based on the IRAS catalogue with a 2 arcminute beam. Radio continuum observations strong enough to be used for astrometry \citep{Morgan90} also  identify the radio continuum source within $<$ 1$"$ of our position. 

The visibilities shown in Figure \ref{Fig:uvampL1641} confirm that the envelope is resolved and that the structure variations are extend inwards to 60 k$\lambda$, a significant difference with NGC 2071. No unresolved component dominates the emission between 20 and 60 k$\lambda$. Any unresolved component that could correspond to a central disk is limited to $<$0.1 Jy and $<$400 AU in size. The envelope structure, as seen by resolved out emission at shorter baselines, extends all the way inwards to 700 AU. 

\begin{figure*}
\begin{center}
\includegraphics[width=400pt]{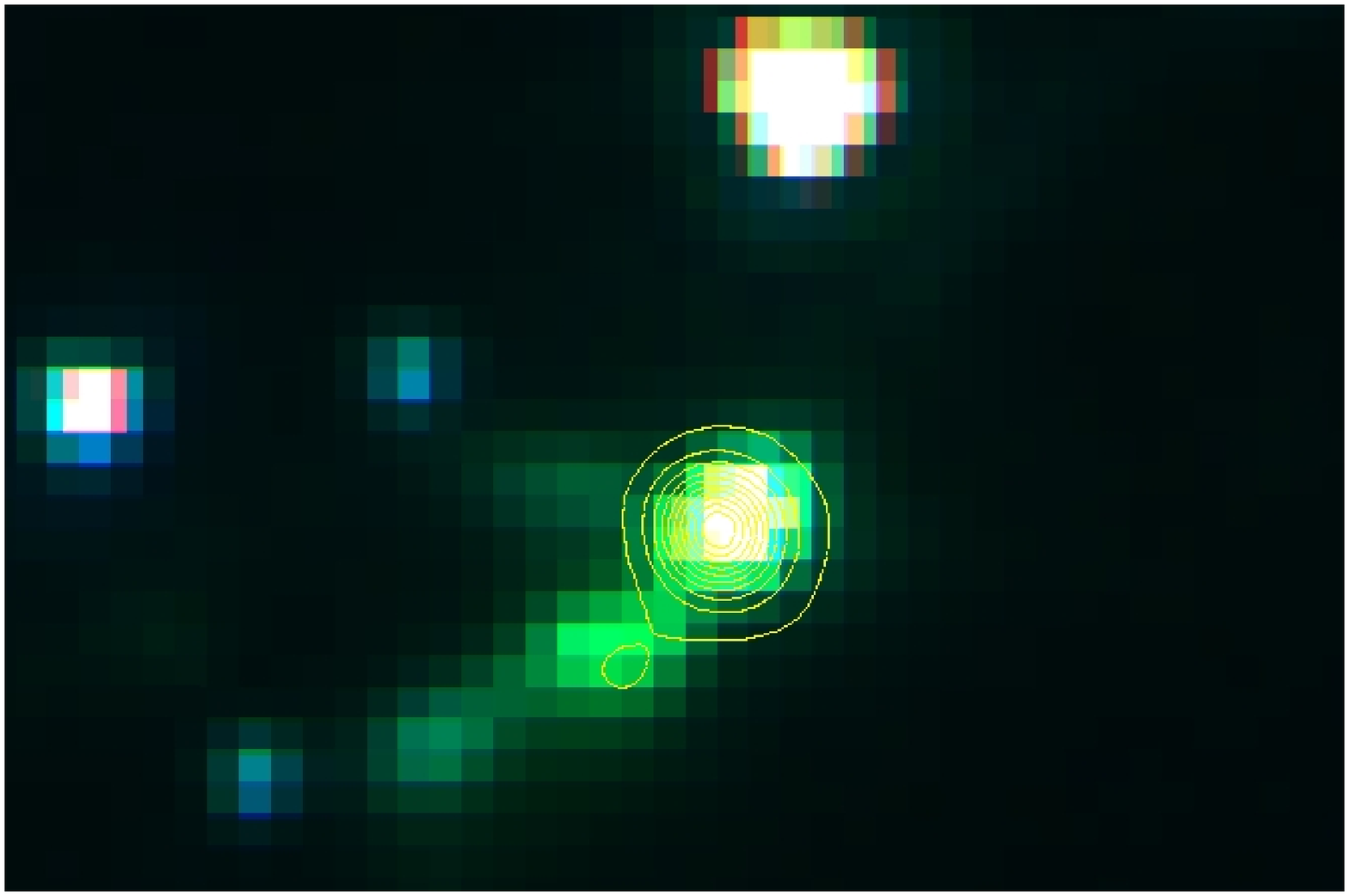}
\end{center}
\caption{SMA continuum observations at 230 GHz at 3,6,9... $\sigma$  of the compact configuration  Fig. \ref{fig:1:cont}  plotted in contours with $\sigma$ = 2.3 mJy/beam in over a Spitzer IRAC 3.6, 4.5 and 8 $\mu$m false color image of L1641 S3 MMS1 }
\label{Spitz:1641}
\end{figure*}

The Spitzer image for L1641 S3 MMS 1, Figure \ref{Spitz:1641}, shows that the embedded source is associated with the bright greenest emission (4.5 $\mu$m). The 4.5 micron IRAC band contains spectral features which emit strongly in shocked molecular gas \citep{DeBuizer10}, so excess emission in this band is often attributed to outflows from YSOs \citep{Cyganowski08,Chambers09}. Other potential members in the L1641 protocluster can be seen, but none emit at 1.3 mm wavelengths. Megeath et al. (in prep) classifies these other sources as normal stars or TTauri stars.  The flow was earlier also seen at \HH emission \citep{Stanke00}.

\section{Analysis}

\subsection{Mass derivations}
If one assumes that the cold dust has the emission properties proposed by column 5 of the table \citet{Ossenkopf94}, also known as OH5 dust, one can calculate the observed mass at each physical scale shown in Figure \ref{fig:1:cont}. The mass can be calculated directly from the continuum emission using the following formula \citep[see e.g.][]{Shirley00};
\begin{equation}
M_D = S_\nu D^2 / B_\nu(T_{dust}) \kappa_\nu,
\end{equation}
where $S_\nu$ is the integrated flux density, $B_\nu(T_{\rm{dust}})$ is the Planck function at the dust temperature $T_{\rm{dust}}$, and $\kappa_\nu$ is the opacity per gram of gas and dust. It is assumed that at both observed wavelengths (1.3 mm and 850 $\mu$m) the dust is optically thin.  \footnote{Note that this may be a source of error. E.g. \citet{Andrews05} found that about 25$\%$ of the emission of dust may be optically thick. It follows that even at 1.3 mm, there is a sizeable contribution of the optically thick emission. However, for simplicity we continue with the assumption that the emission is optically thin. } For OH5 dust, $\kappa_\nu$ is  0.02 cm$^2$ gm$^{-1}$ at 850 $\mu$m and 0.009 cm$^2$ gm$^{-1}$ at 1.3 mm. Without additional measurements, however, the dust temperature cannot be independently derived. At large scales, the mass is dominated by the outer cold dust and the temperature is typically assumed to be 20 K \citep{Shirley00} and  we adopt that value in this paper. Material heated to higher temperatures, e.g.  by outflow shocks,  will also emit at these wavelengths but should not account for a large fraction of the mass. Thus, we assume that the bulk of the material is dominated by the cold gas on all spatial scales. Table \ref{tab:masses} provides the derived masses as a function of spatial scale.
  
The derived total mass is likely to have a significant level of uncertainty due to uncertainties in the observed fluxes, the gas to dust ratio,  and the underlying range of possible dust properties. We estimate the uncertainty in the total mass at about 40$\%$ from these effects and refer the reader to \citet{Shirley00}, \citet{Shirley02} and  \citet{diFrancesco08}  for a more in-depth discussion. The uncertainty in the dust temperature, however, can introduce a significant uncertainty in the derived mass. Using a lower dust temperature of 10 K increases the mass by a factor of about three, while assuming a higher temperature of 30 K yields about 2.5 times less mass.

\placeTablemassresults

On large scales the masses are dominated by the cold dusty envelopes. Only small fractions ($<$0.5 \Msol) remain visible at the highest resolutions observed with the eSMA. Even if a large amount, e.g. 75$\%$, of the disk material is resolved out by the lack of shorter baselines, disk masses would still be below 2 \Msol. 
A scenario in which the disks are largely resolved out is unlikely, since at the distance of Orion, $\sim$450 pc, the resolved eSMA disks are still $\sim$200 AU in radius. Within the extended observations of L1641 S3 MMS1, which probes down to scales of 400 AU, the mass ($\sim$3.8 \Msol) is still dominated by the cold outer envelope, extending down to a few hundred AU. An unresolved disk( $<$ 400 AU) would have a mass  $<$0.45 \Msol, assuming the unresolved inner part has  brightness $\sim$0.2 Jy as seen from the longest baselines. In conclusion, all sources, independent of possible IR detections with Spitzer are clearly deeply embedded sources equivalent to the Stage 0. 

\subsection{Spectral Energy Distributions}

\begin{figure*}
\label{fig:rob}
\begin{center}
\includegraphics[width=200pt]{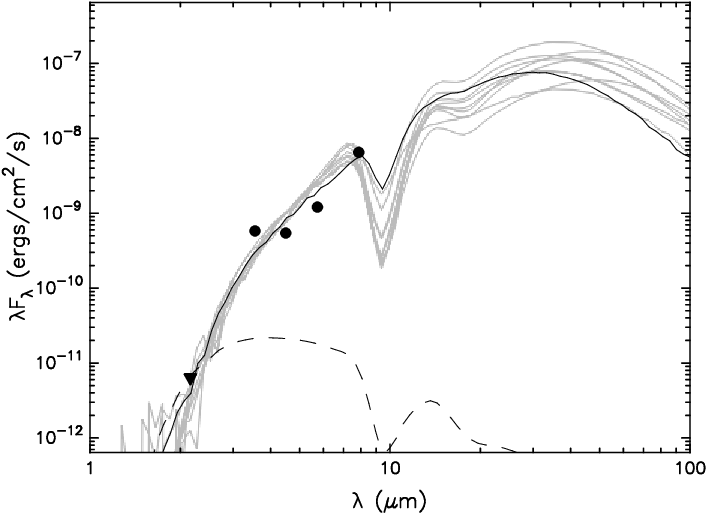}
\includegraphics[width=200pt]{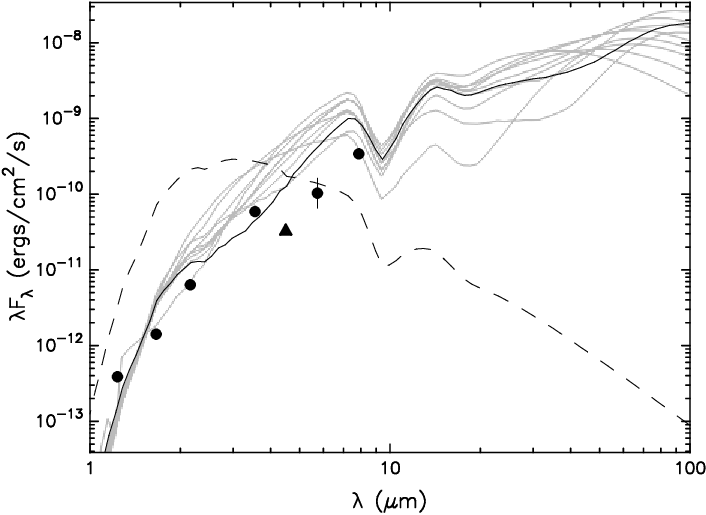}
\includegraphics[width=200pt]{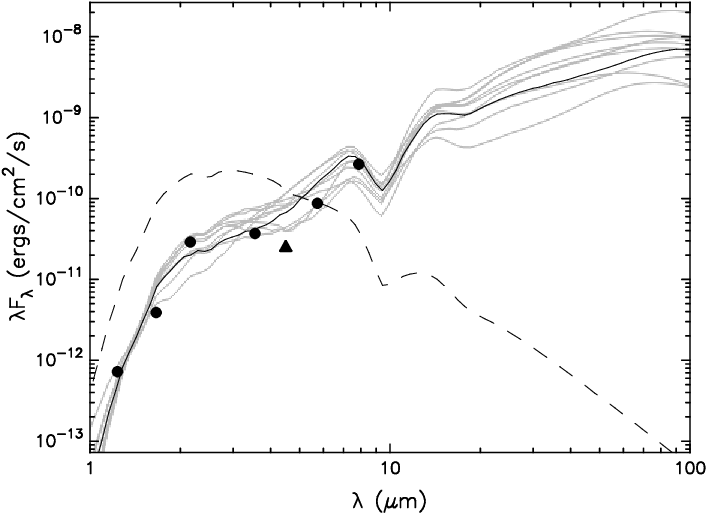}
\includegraphics[width=200pt]{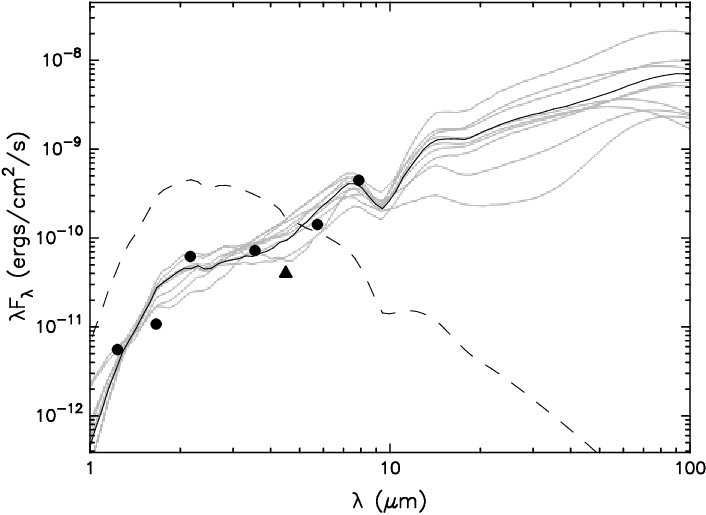}

\end{center}
\caption{SED fits  for ({\it Top Left}) L1641 S3 MMS 1, ({\it Top Right}) 2071-A, ({\it Bottom Left}) 2071-B, and ({\it Bottom Right}) 2071-C. The black line shows the best-fitting model, the grey lines show the next 10 best fits. Solid circles show confirmed fluxes, while triangles show either lower (normal triangle) or upper limits (upside-down triangle). Bars show the uncertainty in the flux determination. The dashed line is the stellar SED if it would be visible unobstructed by circumstellar material.}
\end{figure*}

A powerful probe of protostellar evolution is the Spectral Energy Distribution (SED). Studies, such as demonstrated in e.g. \citet{Robitaille07} and \citet{Evans09}, have used databases of 2MASS and Spitzer sources to efficiently  characterize large number of protostars using their SEDs. Although some studies, such as \citet{Crapsi08}, \citet{Enoch08}, \citet{vanKempen09},  and Maxwell $\&$ Johnstone (in prep), have shown that individual SEDs may not produce unique classification and physical structure parameters, they nevertheless do provide a reasonable guess at the evolutionary stage and structure of the central protostar.
Using the Spitzer data, complemented by 2MASS data where available, SEDs in the range of 1 to 100 $\mu$m were constructed for all four detected sources (2071-A,B and C and L1641 S3 MMS 1) and in turn fitted with a model  from the SED grid of models presented by \citet{Robitaille06} using the online SED 
fitter\footnote{The SED fitter can be found at http://caravan.astro.wisc.edu/protostars/} \citep{Robitaille07}.  The submillimeter observations were not included due to either their uncertain calibration (SCUBA) or the lack of short spacings and the filtering out of emission at larger scales (SMA). 

Due to the possible inaccuracy of individual SED fits shown by \citet{Crapsi08}, as well as the duplicity of many fits \citep{Robitaille06,Robitaille07}, we average the 10 best SED results and study only the emission shortwards of 100 \micron (models which grossly under- or overestimate the long wavelength fluxes are dismissed before the averaging). Figure \ref{fig:rob} shows the fitted SEDs for L1641 S3 MMS1, 2071-A, 2071-B and 2071-C. Table \ref{tab:SED} in turn presents the resulting envelope masses, bolometric luminosities, stellar masses and the ratio of envelope over stellar mass of the SED, derived by averaging the 10 best fits. \\
\placeTableSED
Table \ref{tab:SED} reveals that the combined luminosities of the three NGC 2071 sources do not account for the previously derived bolometric luminosity of 520, assumed for the total core. Most of the high core luminosity on larger scales is due to the inclusion of IRAS 60 and 100 \micron photometry, which includes a significant contribution from the ISM and cloud material surrounding the core. Even the removal of these points indicates that the bolometric luminosity is dominated by large-scale cold cloud material. Higher resolution imaging using the PACS and SPIRE instruments on Herschel, as proposed by HOPS (PI: T. Megeath) will re-observe these regions and determine more accurate core luminosities. 

The envelope masses found via SED fitting and direct measurement of dust continuum emission range from a few solar masses for 2071-C to almost 15 \Msol for 2071-B. This is significantly higher than the masses commonly found in nearby clouds, which are typically a few tenths to $\sim$2 \Msol \citep{Shirley00,Johnstone00,Johnstone01}.
However, the masses derived from the SED are within 20 $\%$ of the masses derived from the compact SMA emission at 230 GHz. 

  Although the Robitaille SED fitter can be used to determine limits on the disk emission, the high \Av and increasing optical depth of the dust makes this very uncertain and untrustworthy for these observations.  We do, however, derive estimates on the stellar masses. From the SED fitting, the L1641 S3 MMS 1 internal source is significantly more massive (3.5 \Msol) than any of those in the NGC 2071 minicluster ($<$ 1 \Msol) and the ratio of envelope over stellar mass is much lower (this is also true if the SCUBA measurements are used to determine the envelope mass). This may indicate that L1641 S3 MMS 1 has had a different accretion history than the NGC 2071 cluster members. Either it is older, or the accretion rate was significantly higher.  Changes in the accretion history produce a significant influence on the radiation field produced by the star at UV wavelengths. We will be investigating this effect using molecular line spectroscopy in a forthcoming paper. 

%
%
%

\section{Discussion on fragmentation}
Although a few individual IM sources have been observed and analyzed in detail (e.g. NGC 7129 IRS 2,
\citet{Fuente05a,Fuente07}, IRAS 20050+2720, \citet{Beltran08}), it is uncertain if these are (i) typical of their 
evolutionary stage or age, (ii) typical of clouds forming more massive stars than those in nearby star-forming 
regions, e.g. Taurus or Ophiuchus (iii) typical for fragmentation at small ($\sim$ 500 AU) scales. Ideally, a much
larger sample of IM sources needs to be considered, and these objects must be investigated at higher spatial
resolution.

The interferometric observations of NGC 2071 and L1641 S3 MMS 1 allow for an investigation into their fragmentation 
histories. Although similar on large scales, these two IM sources are quite different on small scales, with one separating
into multiple components and the other remaining a single protostar. In this section we will attempt to unravel the 
fragmentation history of two cores massive enough to form an intermediate mass protostar and determine which physical
parameters make this possible. For an extensive in-depth discussion of fragmentation, we refer the reader to the series of papers by 
Alan Boss \citep[e.g.][]{Boss93,Boss02,Boss09} and Matthew Bate \citep[e.g.][]{Bate98,Bate05,Bate09}.

\subsection{Conceptual Model}

To make the problem of molecular cloud fragmentation tractable, we start by assuming the intrinsically continuous fragmentation 
process of an original core can be separated into N discrete time steps: t=t$_0$, t$_1$, t$_2$,... ending with what is observed today 
at t=t$_{\rm{now}}$. Assuming the gas is of uniform density, isothermal and thermally supported against gravitational contraction, 
the mass and separation of gravitationally induced fragments is given by the Jeans equations, written in terms of density and temperature.

\begin{equation}
{M}_J = 1 \left(\frac{T}{10\,{\rm K}}\right)^{\frac{3}{2}} \left(\frac{n}{10^4\, {\rm cm}^{-3}}\right)^{-\frac{1}{2}}\, M_{\odot}
\end{equation}
and
\begin{equation}
 \Lambda_J = 2\times 10^4  \left(\frac{T}{10\, {\rm K}}\right)^{\frac{1}{2}} \left(\frac{n}{10^4\, {\rm cm}^{-3}}\right)^{-\frac{1}{2}}\,\rm{AU}
\end{equation}

It is important to note that both $M_J$ and $\Lambda_J$ are minimum requirements for gravitationally induced fragmentation and that
for spherical condensations without inhomogeneities the fastest growing collapse mode consists of the entire core. For non-spherical
condensations, or cores with significant inhomogeneities the opportunity for fragmentation at these Jeans scales is much enhanced
\citep[see for example][]{Pon11}.

The density, $n$, and temperature, $T$, of the gas will change with time due to the collapse itself, as well as the additional infall of large-scale
material. The average density and temperature at each time-step can be denoted $T_0$ and $n_0$, $T_1$ and $n_1$, ... Depending on the 
local physical conditions at a given time the cloud may break up into ever smaller fragments. We denote the observed fragments as $A, B, C$, ....
and the precursor as the sum of final fragments (i.e. for three observed fragments $A$, $B$, and $C$, the initial core would be designated ABC).

From the given parameters, one can derive two basic scenarios that a cloud can follow during fragmentation : (1) Direct fragmentation: A cloud directly fragments into the resulting structure in a single time-step.  In this scenario the conditions for fragmentation are reached throughout the cloud and the resulting cores are all co-eval. There is only a single event t$_1$. \textit{Notation: ABC $\Rightarrow$ A+B+C}; (2) Hierarchical fragmentation.  In this scenario it is 
possible for a fragment created in time-step t$_x$ to fragment again into
smaller fragments at a later time steps, which we label t$_{x1}$, t$_{x2}$,.... This can be done as long as the requirements as defined by the Jean's equations are met. \textit{Notation: ABC $\Rightarrow$ AB+ C or AC+B or A+BC $\Rightarrow$ A+B+C}.

A fragmentation event is characterized by two numbers, $\Lambda_J$ and $M_J$.  The number of possible fragments created can be determined by 
dividing the mass of the cloud by $M_J$, however, as noted above the actual degree of fragmentation is determined by inhomogeneities in density and 
geometry (assuming for the moment a cloud held up only by thermal support and thus neglecting additional support mechanisms such as rotation or magnetic fields).
The final mass of a fragment is likely to be much higher than the instantaneous Jeans mass, both due to this under-fragmentation effect and the addition of further material from the accreting core. The fragmentation scale, however, is expected to provide a more durable result - close fragments must form at high densities and low temperatures.

 \subsection{Applying the conceptual model to NGC2071}
        
Returning to the case of NGC 2071, we observe three present-day key components: A, B and C (see Fig 5). Although a tentative detection of a fourth source, VLA-1 (or D), is discussed above, the lack of emission at infra-red wavelengths uniquely tied to this position means we cannot rule out this being a shock position. The application of the conceptual model for three or four fragments is identical, including the caveats discussed later. We assume these formed from the
same initial core, ABC, at t=t$_0$. This means there can only be either one or two fragmentation steps: either i) ABC directly fragmented into
A, B and C, or ii) ABC first fragmented in two (AB+C, AC+B or A+BC), and one of these subsequently fragmented. 
We further assume that the separation of the cores, $\Lambda_{\rm{XY}}$, has not changed 
significantly between the fragmentation event and the time of observation: t$_{\rm{obs}}$.

We first note that the fragments appear to have a hierarchical structure, with fragments A and B, near the core center, with a small separation $\Lambda_{AB} \sim 2500\,$AU and fragment C, closer to the core edge, with a separation $\Lambda_{(AB)C} \sim 5000\,$AU from the AB
pair.\footnote{The formation and fragmentation of a mini-cluster is formed in three dimensions and not in two.  As such the observed distances, 
and the corresponding Jeans lengths, $\Lambda$,  are projected distances and not  physical distances. The difference in projected and physical distance 
between AB, BC and AC can of course be different. If one assumes there is a significant depth in the core , distances can be assumed to be larger by a 
factor 1/$sin(i)$ with $i$ the angle between two cores along the line of sight. On average this will be 30$\%$ and at this level of uncertainty these results still hold.}
Considering the Jeans equations above,
the simplest way to account for the separation variation is to assume that the density was higher by a factor of four (or the temperature was lower by a factor of four) when A and B fragmented
compared with C. Similarly, the best handle on the density at fragmentation is supplied by the separation of the fragments\footnote{It is of course possible that there is a dynamical component that changes the location of either A, B or C after formation. From unpublished data we observe a small difference of $<$0.15 km s$^{-1}$ between C and AB, while there is no detectable difference $<$0.1 km s$^{-1}$ between A and B in line observations of C$^{18}$O (van Kempen et al. in prep). This limits the movement of C as compared to AB to $<$3,000 AU, assuming a lifetime of 0.08 MYr (50$\%$ of the average Class 0 lifetime \citep{Evans09}). It is therefor not possible to rule out that Core C was ejected by A and B, although it is unlikely given the calibration errors in the velocity determination.}, as noted in the previous section, and thus we
suggest that the fragments formed when the gas density was at least $3 \times 10^5$ cm$^{-3}$ (assuming an isothermal gas at $T=20\,$K). For these conditions, the
instantaneous Jeans mass of each fragment would only be $\sim 0.5\,M_\odot$, suggesting that additional accretion of material onto each fragment was required.
\placeFigurefragment
To reproduce the higher masses of the central fragments A and B, compared with fragment C, utilizing only direct fragmentation requires either lower initial densities or higher temperatures. Conversely, the smaller separation between the fragments, and their proximity near the center of the core, suggest higher density and lower temperature conditions.
The simplest explanation for the mass discrepancy is that the original core did not fully fragment into $M/M_J$ pieces (where $M_J \sim 0.5\,M_\odot$), and thus the masses of A and B do not reflect the instantaneous Jeans mass but rather reflect the bulk mass of the inner core after further accretion. Note it is possible D is one such fragment, but that due to its location it was deprived of incoming mass by A and B. Under such a scenario, the original core would have had only mild inhomogeneities, resulting in only the few fragments observed. One of these fragments, relatively far from the core center accreted little material beyond its initial fragment whereas the two central fragments continued to accrete the bulk of the core material.

We thus propose the following cartoon history for NGC 2071, as shown in Fig. \ref{fig:frag}. The original supercore, ABC, collapsed isothermally until reaching a critical density 
$n_1 > 3 \times 10^{5}$ cm$^{-3}$ in the central region ($\sim$ 5,000 AU) at $t_1$ where it fragmented into the cores AB and C  with the observed separation.  At $t_{2}$ the mean density had increased by a factor of four and AB fragmented into A and B.  After $t_{2}$, A, B and C formed protostars, accreting further mass from the natal supercore. The heating of these central stars inhibited further fragmentation due to the increased gas temperature, since the Jeans' lengths of remnant density irregularities increase to scales larger than the core. During all this the supercore may well have continued to accrete from the global mass reservoir seen at larger scales.

Whether the core fragmented hierarchically in time is debatable. It is likely that the core initially fragmented into AB and C with a later fragmentation of 
AB into A and B when the central density had increased significantly. However, observed pre-stellar cores are known to have increasing density toward their 
centers (and decreasing temperatures \citep{Bergin06}) and thus a single fragmentation scenario cannot be ruled out. 

In the above analysis we have inherently assumed  that the thermal pressure is dominant. However, based on single dish measurements and our SMA molecular line data (van Kempen et al. in prep.), we know that cores are highly supersonic (> 5). Clearly the cores are not thermally supported. Including such a significant non-thermal motion as an equivalent pressure in eqn. (2) and (3) of Jeans analysis has been shown to be more consistent with the observed core masses and separations \citep[e.g.][]{Pillai11}. From our SMA C$^{18}$O data, the line FWHM is ~2.5 km/s, corresponding to a velocity dispersion of 1.1 km/s.
For a density of 10$^6$ cm$^{-3}$ following eqn. (3) and (4) of \citet{Pillai11}, we then derive a  Jeans mass  of 100 \Msol, much higher than the total mass of the clump as derived from archival SCUBA data.  The corresponding Jeans length also would require the cores to be separated by $>$20,000 AU inconsistent with the observed projected separations. However increasing the density to few times 10$^7$ cm$^{-3}$ would give Jeans mass and length consistent with the observed values. Such initial core densities are extreme (and unlikely) and therefore, we conclude that  turbulent jeans fragmentation to present values would not justify our observations.

Perhaps the most interesting aspect of this analysis is the suggestion that the NGC2071 core underwent much less fragmentation than could have been possible given a simple Jeans stability argument. The present core density implies
that the number of possible fragments is in the hundreds and even assuming somewhat lower initial density
conditions, the number of fragments could have numbered more than ten.   This suggests that there may be additional support mechanisms working against gravitational fragmentation. Examples of support mechanisms include: rotation, turbulence, and magnetic fields. We will discuss these in the next section, as these support mechanisms are also the reason L1641 S3 MMS 1 did not fragment at all. Even with the inclusion of VLA-1 as a primordial fragment, these support mechanisms would still have prevented fragmentation at similar level.


\subsection{Why did L1641 not fragment?}

Given the large-scale similarities between NGC 2071 and L1641 S3 MMS 1 it is curious that the later did not fragment. However, as noted at the end of the 
previous  section, even NGC2071 seems to have undergone less fragmentation than might be expected from a Jeans gravitational instability argument. From this
perspective it would seem that NGC 2071 and L1641 S3 MMS1 are more similar than they are different and that in both cases support mechanisms against collapse 
are required. One additional parameter can also be raised for L1641 S3 MMS 1 and that is the lack of inhomogeneity. 

We note that for low mass pre-stellar cores very little initial inhomogeneity has been observed in the density structure beyond that required to support against gravity \citep{Schnee10}, suggesting that these objects will at least begin their collapse to protostars monolithically. The evolution of isothermal structures with constant density depends explicitly on the detailed geometry as global, large-scale collapse, competes against gravitational fragmentation \citep[See ][ for a discussion]{Pon11} In general the more spherical the initial core, the less fragmentation expected.
\subsubsection{Support Mechanisms against fragmentation}
Additional support mechanisms against fragmentation which should be considered include rotation, turbulence, and magnetic fields. For example,  3D hydrodynamical simulations by \citet{Bate98} show that sufficient rotational energy in a cloud or core prevents fragmentation as the growth of the non-axisymmetric perturbations is stifled. 
Gravitational torques actually remove the angular momentum. Close binaries can still be formed. Indeed \citet{Carrasco12} reveals that 2071-C is likely a close binary ($\sim$500 AU). 

  Turbulence has a similar stifling effect \citep{Bate09} by preventing or delaying the growth of non-axisymmetric perturbations. Turbulence can be introduced by e.g. ambipolar diffusion of external radiation sources or the formation of the first protostar and associated outflow. 
  The inclusion of magnetic fields highly complicates the formation and fragmentation of protostellar cores in molecular clouds. Similar to turbulence and rotation, magnetic fields are able dampen the fragmentation through the process of magnetic braking as well as magnetic pressure \citep{Hennebelle08,Price07}.
But, magnetic fields can also enhance fragmentation \citep{Boss09}. Magnetic tension prevents accretion and thus the fast growth of a central density singularity. This tension drives small density variations toward the Jeans' mass.  A key parameter that determines the scale of the magnetic tension and raking seems to be the  initial shape of the molecular cloud with respect to the direction of the magnetic field. For instance, an oblate cloud will fragment much more than a prolate cloud. The study of fragmentation of detailed MHD codes is a very active field with many parameters that need to be included. A main conclusion is that the effect of magnetic fields is highly dependent on the initial conditions, in particular the small-scale structure and the inclusion of all physical characteristics, e.g. ambipolar diffusion.  
%


\section{Conclusions}

In this paper we have presented millimeter and submillimeter interferometry observations of two  intermediate mass (IM) protostellar cores in Orion, L1641 S3 MMS 1 and NGC 2071. The following conclusions can be drawn from the data:
\begin{itemize}

\item Continuum observations from the JCMT/SCUBA and the SMA reveal that at large scales these two sources are similar in mass ($\sim$25 \Msol) but that at smaller spatial scales NGC 2071 has fragmented into three, or potentially four, low mass (LM) protostars, which is not observed in L1641 S3 MMS1. Central stellar masses derived from SED fitting reveal that NGC 2071 contains low-mass sources, with stellar masses $<$ 1 \Msol. The central star in L1641 S3 MMS 1 is $>$ 3 \Msol. Although it is possible the protostars in NGC 2071 will still form A-stars, only L1641 S3 MMS1 will likely form a B7-B9 star.  All three protostars in NGC 2071 remain deeply embedded, based on the  the ratio of envelope over stellar mass, and thus Class 0 or Stage 0 sources as defined by the classification of \citet{Evans09}. L1641 S3 MMS1 is a much more massive protostar (stellar mass $\sim$ 3 \Msol) and has not fragmented. It also classified as Stage 0, despite being bright in the infrared (Spitzer).  
\item Disks around the three sources in NGC 2071 are constrained to masses smaller than 0.35 \Msol, about 2$\%$ of the total mass of the envelopes and 10$\%$ of the derived stellar masses. The possible disk around L1641 S3 MMS1 cannot be fully characterized due to a lack of very long baselines, but is also constrained to $<$ 0.45 \Msol, $<$ 4 $\%$ of the total mass.
\item Calculation of Jean's masses and lengths and subsequent comparison to the physical distribution shows that much less accretion must have taken place on the NGC 2071 cluster as well as on L1641 S3 MMS 1. In the latter case the stifling prevented fragmentation altogether. The stifling likely was caused by a combination of support mechanisms against fragmentation, such as rotation, turbulence and magnetic breaking. Fragmentation at very small scales ($<$ 100 AU) cannot be ruled out. 
\item The bulk of accretion onto the three NGC 2071 cores took place {\it after} the fragmentation. These results favor a model of competitive global accretion from a large-scale molecular cloud. Fragmentation might thus have taken place when each core was $\sim$0.5 \Msol, about 10$\%$ of their current masses. 
\end{itemize}

Future studies of these sources should focus on observations using even longer baselines and thus higher spatial resolution. Studies with e.g. the Atacama Large Millimeter/Submillimeter Array (ALMA) can probe down to resolution of 0.2-0.1 $"$. These will be able  accurately characterize the protostellar disks surrounding the stars at high precision.

\acknowledgments
TvK and SL were supported as SMA postdoctoral fellows at the Harvard-Smithsonian Center for Astrophysics(CfA) and are thankful for the Submillimeter Array for funding their research. TvK's current research is supported by NOVA (Nederlandse Onderzoeksschool Voor Astronomie). TvK is also grateful for the facilities at the Joint ALMA Observatories during his association. DJ acknowledges the support from an NSERC Discovery Grant. TP acknowledges support from the Combined Array for Research in Millimeter-wave Astronomy (CARMA), which is supported by the National Science Foundation through grant AST 05-40399. The anonymous referee is thanked for the critical useful positive read-through of the paper. Mark Gurwell, Ken Young (Taco) and David Wilner of the CfA are thanked for track-sharing eSMA observations on short notice. Taco and Remo Tilanus are thanked for their efforts on the eSMA and assistance with data reduction. We are grateful for the other members in the WISH Intermediate mass team (Lars Kristensen, Mike Fich and Carolyn McCoey in particular) for supplying necessary information on the sources and  general discussions. Last but not least, we would like to express our appreciation for the help of Tom Megeath by sharing the reduced Spitzer photometry and images on Orion before publication.

{\it Facilities:} \facility{SubMillimeter Array, Spitzer Space Telescope, eSMA}

\bibliographystyle{aa}
\bibliography{biblio}

\end{document}